\documentclass[10pt,aps,prx,twocolumn,superscriptaddress]{revtex4-2}
\usepackage{graphicx,amsmath,amsfonts,amssymb,mathtools,bm,times}
\usepackage{relsize,epstopdf,upgreek,color,xcolor,rotating,tabularx,booktabs,setspace,makecell}
\usepackage{cancel}
\usepackage[colorlinks=true,linkcolor=blue,citecolor=blue,urlcolor=blue]{hyperref}

\newcommand{\argp}[1]{\left(#1\right)}
\newcommand{\args}[1]{\left[#1\right]}
\newcommand{\argc}[1]{\left\{#1\right\}}
\newcommand{\ket}[1]{\left| #1\right>}
\newcommand{\bra}[1]{\left< #1\right|}
\newcommand{\ketbra}[2]{\ket{#1}\!\!\bra{#2}}
\newcommand{\inner}[2]{\langle #1|#2\rangle}
\newcommand{\opinner}[3]{\langle #1|#2|#3\rangle}
\newcommand{\rvec}[1]{\pmb{#1}}
\newcommand{\tr}[1]{\mathrm{tr}\!\left\{#1\right\}}
\newcommand{\ptr}[2]{\mathrm{tr}_{#1}\left\{#2\right\}}
\newcommand{\Tr}[1]{\mathrm{Tr}\{#1\}}
\newcommand{\D}{\mathrm{d}}
\newcommand{\I}{\mathrm{i}}
\newcommand{\E}[1]{\mathrm{e}^{\mbox{\footnotesize$#1$}}}
\newcommand{\MEAN}[1]{\left<{#1}\right>}
\newcommand{\AdA}{a^\dagger a}
\newcommand{\AVG}[2]{\underset{#1}{\displaystyle{\mathop{\mathbb{E}}}}\!\left[#2\right]}
\newcommand{\VEC}[1]
{|#1\rangle\negmedspace\rangle}
\newcommand{\BVEC}[1]
{\langle\negmedspace\langle#1|}
\newcommand{\etad}{\eta^\prime}
\newcommand{\Us}{U_\mathrm{s}}
\newcommand{\khat}{\widehat{\mathbf{k}}}
\newcommand{\nhat}{\widehat{\mathbf{n}}}
\newcommand{\SL}{C_\textsc{l}}
\newcommand{\rhoq}{\rho_\textsc{q}}
\newcommand{\rhob}{\rho_\textsc{b}}

\begin{document}
	
	\author{Saurabh U. Shringarpure}
	\affiliation{NextQuantum Innovation Research Center, Department of Physics and Astronomy,  Seoul National University, Seoul 08826, South Korea}
	
	\author{Siheon Park}
	\affiliation{NextQuantum Innovation Research Center, Department of Physics and Astronomy,  Seoul National University, Seoul 08826, South Korea}
	
	\author{Sungjoo Cho}
	\affiliation{NextQuantum Innovation Research Center, Department of Physics and Astronomy,  Seoul National University, Seoul 08826, South Korea}
	
	\author{Yong Siah~Teo}
	\email{ys\_teo@sejong.ac.kr}
	\affiliation{Department of Quantum Information Science and Engineering, Sejong University, Seoul 05006, South Korea}
	\affiliation{NextQuantum Innovation Research Center, Department of Physics and Astronomy,  Seoul National University, Seoul 08826, South Korea}
	
	\author{Hyukjoon Kwon}
	\email{hjkwon@kias.re.kr}
	\affiliation{School of Computational Sciences, Korea Institute for Advanced Study, Seoul 02455, South Korea}
	
	\author{Srikrishna Omkar}
	\email{omkar.shrm@gmail.com}
	\affiliation{NextQuantum Innovation Research Center, Department of Physics and Astronomy,  Seoul National University, Seoul 08826, South Korea}
	
	\author{Hyunseok Jeong}
	\email{h.jeong37@gmail.com}
	\affiliation{NextQuantum Innovation Research Center, Department of Physics and Astronomy,  Seoul National University, Seoul 08826, South Korea}
	
	\title{Code-agnostic bosonic noise suppression with hybrid rotations}
	
	\begin{abstract}
		Physical-level noise on traveling bosonic modes remains a critical bottleneck for scalable quantum information processing. We show that for any single-mode bosonic code (qumode) corrupted by thermal or Gaussian displacement noise at loss rate $\mu$ and amplification $G$, a hybrid continuous-discrete-variable (CV-DV) interferometer using a single qubit ancilla and two controlled-Fourier (CF) gates sandwiching the noise channel suppresses its effects from linear to quadratic scaling. This is achieved \emph{without} active error correction or destructive measurements of the encoded state, maintaining high success probabilities $\geq 0.5$ when $\mu G \leq 0.5$. When supplemented with multiple ancillas, the protocol converts photon loss into coherent Fock-damping, and thermal or displacement noise into a mixture of Fock-diagonal noise. The protocol is entirely code-agnostic. For the special case of $2^K$-fold rotation-symmetric bosonic codes, it simplifies to conventional error detection and projection with $K$ ancillas. Suppression with simple gates and few ancillas demonstrates a clear hardware-efficient advantage over previously proposed ``bypass'' schemes, where quantum information transferred to the DV ancillas is readily corrupted by ancilla noise. Finally, we extend the protocol to a qutrit DV ancilla. This demonstrates resilience to both CV noise and composite DV damping noise, achieving a truly hybrid noise suppression scheme that operates effectively even on CV encodings lacking a well-defined photon-number \mbox{parity syndrome}.
	\end{abstract}
	
	\maketitle
	
	\section{Introduction}
	
	Controlled interactions between discrete-variable (DV) and continuous-variable (CV) quantum systems are foundational to architectures including cavity~\cite{Haroche1989:Cavity, Raimond2001:Manipulating} and circuit quantum electrodynamics~\cite{Wallraff2004:Strong, Blais2004:Cavity, Heeres2015:Cavity, Krastanov2015:Universal, Blais2021:Circuit}, and trapped ions~\cite{Leibfried2003:Quantum, Lv2018:Quantum}. 
	Among these, bosonic modes (qumodes) and their hybrid circuits~\cite{Xiang2013:Hybrid, Andersen2015:Hybrid, Albert2018:Performance, Joshi2021:Quantum} have emerged as particularly promising platforms for quantum information processing.
	
	While stationary bosonic encodings inside superconducting cavities are well established~\cite{Leghtas2013:Hardware-Efficient, Leghtas2013:Deterministic, Ofek2016:Extending, Sivak2023:Real, Ni2023:Beating, Campagne2020:Quantum, Eickbusch2022:Fast, Gao2021:Practical}, traveling qumodes are essential for scalable, distributed quantum computing through long-distance interconnects~\cite{Kimble2008:Quantum, Reiserer2015:Controlled, Hacker2019:Deterministic}, since they are the natural carriers of quantum information between physically separated nodes. Hybrid CV-DV systems offer a natural route to hardware-efficient fault tolerance~\cite{Omkar2021:Highly, Xu2024:Fault-Tolerant, Liu2026:Hybrid, Lee2025:Photonic}, including in trapped-ion platforms where vibrational modes serve as bosonic qumodes~\cite{Kang2025:Doubling}. However, noise critically limits these architectures by drastically increasing resource overheads for downstream error correction~\cite{Omkar2020:Resource-Efficient, Lee2024:Fault-tolerant}, motivating \emph{physical-level noise suppression}.
	
	Photon loss, thermal excitation, and random displacement dominate CV noise. While thermal noise is negligible for optical photons, it remains a hurdle for lower-energy traveling microwaves in superconducting quantum networks~\cite{Kurpiers2018:Deterministic, Axline2018:On-demand}, despite prior theoretical proposals~\cite{Xiang2017:Intracity, Vermersch2017:Quantum} and recent experimental progress~\cite{Qiu2026:Thermal}. Existing suppression strategies present several trade-offs. The ``bypass'' protocols~\cite{Park2022:Slowing} destroy the traveling nature of the qumode by transferring information to stationary DV ancillas. Linear-optical error mitigation~\cite{Taylor2024:Quantum, Teo2025:Linear} incurs high sampling costs~\cite{Takagi2022:Fundamental, Cai2023:Quantum, Quek2024:Exponentially} and prevents further \mbox{\emph{quantum}} information processing. Active squeezing~\cite{LeJeannic2018:Slowing, Brewster2018:Reduced, Schlegel2022:Quantum, Pan2023:Protecting, Rousseau2025:Enhancing} is restricted by the requirement of asymmetric phase-space properties in the encoding or in the noise structure~\cite{Srikanth2008:Squeezed, Omkar2013:Dissipative, Provaznik2025:Adapting}. Beyond losses~\cite{Ralph2011:Quantum, Micuda2012:Noisless}, direct linear-optical suppression of thermal noise that retains the quantum \mbox{state remains challenging}~\cite{Teo2025:Linear}.
	
	In this article, we propose a \emph{code-agnostic}, feedback-free noise suppression scheme for qumodes suffering from photon loss, thermal, or Gaussian displacement noise. The protocol is a hybrid interferometer comprising two controlled-Fourier (CF) gates, each entangling the photon-number parity of the bosonic mode with the DV ancilla, sandwiching the noise channel. Imprinting a parity-dependent phase on each error branch, the CF gate makes photon-number-changing errors acquire mismatched phases on either side of the interferometer, which cancel destructively in the output. The bosonic ladder operators change photon number independently of the encoding, so this destructive interference operates on any single-mode bosonic input state, making the protocol inherently code-agnostic.
	
	The primitives required for an experiment are well-established via the dispersive qubit-cavity interaction in both optical~\cite{Reiserer2014:Quantum, Hacker2019:Deterministic} and superconducting microwave platforms~\cite{Heeres2015:Cavity, Krastanov2015:Universal}. The same conditional-rotation primitive has been analyzed in detail for cat-code repeater schemes~\cite{Li2023:Memoryless, Li2024:Performance}. Our protocol \emph{directly reshapes} the physical noise channel via heralded interactions, an approach previously shown to induce non-unitary CV transformations~\cite{Nodurft2019:Optical}, rather than acting reactively via post-transmission parity measurements~\cite{Sun2014:Tracking, Rosenblum2018:Fault-tolerant}. This fundamental shift eliminates the code-specific constraints that typically restrict parity-based error correction. It generalizes discrete-variable error filtration~\cite{Gisin2005:Error}, where errors are suppressed by quantum interference, to potentially-traveling qumodes, suppressing first-order CV noise effects with a \mbox{single ancilla}. 
	
	For multiple ancillas, we analytically show that photon loss, thermal, and Gaussian displacement noise channels are converted into Fock-diagonal form (diagonal in the photon-number basis), with a finite nonzero success probability reached rapidly in the number of ancillas used. For bosonic codes whose logical codewords share the same photon-number parity, the CF-based suppression is entirely resilient to uncalibrated DV amplitude- and phase-damping noise. This resilience and its multi-ancilla extension stem from a specific class of rotation-symmetric codes for which our protocol reduces to conventional error detection.

	With a qutrit ancilla, the protocol suppresses both CV noise and the composite cascaded amplitude- and phase-damping DV noise. The qutrit extension also covers CV encodings without a well-defined photon-number parity syndrome, which the qubit-ancilla protocol excludes.
	
	The rest of this article is organized as follows. Section~\ref{sec:noise_models} defines the CV and DV noise models used throughout. Section~\ref{sec:single_ancilla} introduces the core hybrid CV-DV suppression interferometer with a single ancilla and demonstrates its code-agnostic nature. Section~\ref{sec:multi_ancilla} extends this to a multi-ancilla protocol, proving rapid convergence toward a Fock-diagonal noise channel. Section~\ref{sec:linear_order_suppression} rigorously derives the quadratic scaling of the suppressed fidelity, and Sec.~\ref{sec:hybrid_entangled} generalizes the protection to hybrid-entangled states. Section~\ref{sec:RSBC} analyzes rotation-symmetric codes and their intrinsic resilience to DV damping noise. Section~\ref{sec:comparison} compares our protocol against complementary strategies of active squeezing, single-ancilla benchmarks built from conditional displacements and rotations, and the multi-ancilla bypass approach of Ref.~\cite{Park2022:Slowing}. Section~\ref{sec:qutrit} presents the qutrit extension and its cascaded DV-noise resilience. Section~\ref{sec:conclusions} summarizes and outlines future challenges.
	
	\begin{figure}[htbp!]
		\centering
		\includegraphics[width=1\columnwidth]{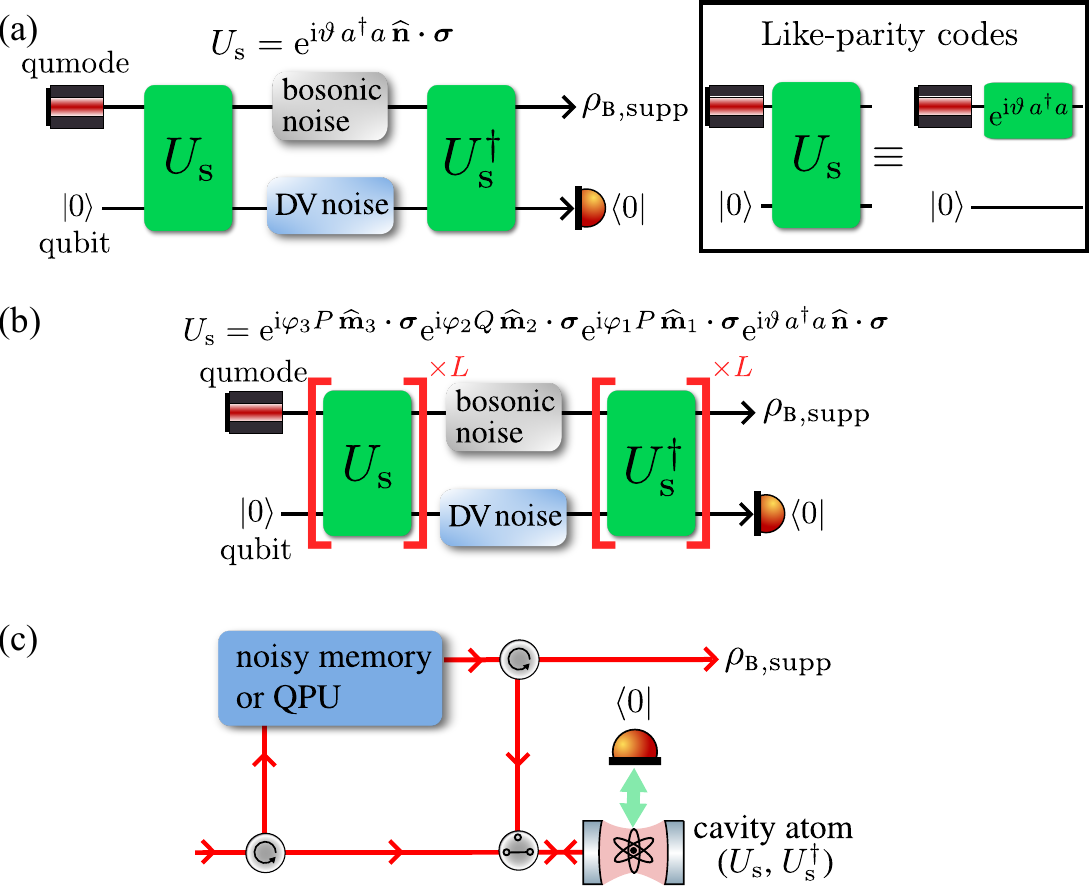}
		\caption{\label{fig:protocols} 
			Bosonic noise suppression with (a) conditional rotation~(condrot) gates and (b) a series of compositions of conditional displacements and the conditional rotation gates. We shall refer to the unitary with $L$ layers by the notation ``[PQP-condrot]$^L$" for convenience. The conditional rotation gate of (a) is optimally the controlled-Fourier gate described in the main text. (c) An optical circuit implementing conditional rotations through the dispersive interactions between a single atom in a cavity and traveling waves. Two circulators route the CV mode around the loop, and a time-dependent optical switch redirects the beams onto a cavity containing an atom to implement the noise suppression unitaries $U_\mathrm{s}$ and $U_\mathrm{s}^\dagger$. The atomic state is read out~(green~arrow) and the CV output $\rho_{\textsc{b},\text{supp}}$ is heralded on measuring the ground state. For bosonic codes with logical states of identical (even) parity, the first hybrid gate can be replaced with a local qumode rotation as shown by the inset in~(a) with $\vartheta=\pi/2$.
		}
	\end{figure}

	\section{Noise models}
	\label{sec:noise_models}
	
	We model CV thermal noise of rate $\eta$ with mean photon number $\bar{n}$, incorporating photon loss as a special case with $\bar{n}=0$, with a quantum-limited amplification channel of gain $G=1+\eta \bar{n}$ following a pure loss channel of rate $\mu=1-(1-\eta)/G$, with $\mu G = \eta(1+\bar{n})$ for this parameterization. The Kraus representations of pure loss and the amplifier are~\cite{Ivan2011:Operator-sum, Gagatsos2017:Bounding, Noh2019:Quantum}, \begin{equation}
		\mathcal{N}_\mathrm{loss}[\rho_\textsc{b}]=\,\sum_{l\geq0}\frac{\mu^l}{l!}(1-\mu)^{\frac{a^\dag a}{2}}\,a^l\,\rho_\textsc{b}\,a^{\dag\,l}\,(1-\mu)^{\frac{a^\dag a}{2}}
	\end{equation} 
	and 
	\begin{equation}  
		\mathcal{N}_\mathrm{amp}[\rho_\textsc{b}]	=\sum_{k\geq0}\frac{(1-G^{-1})^k}{k!\,G}\,a^{\dag k}\,G^{-\frac{a^\dag a}{2}}\,\rho_\textsc{b}\,G^{-\frac{a^\dag a}{2}}\,a^{k},
	\end{equation}
	respectively, where $a$ and $a^\dagger$ are the bosonic ladder operators and $\rho_\textsc{b}$ is the density operator of the bosonic~mode. Gaussian displacement channels decompose similarly into pure loss and quantum-limited amplification (see Appendix~\ref{app:noise-sources} or~\cite{Noh2019:Quantum}), and all derived results in this paper apply to displacement noise as well.
	
	Each DV damping channel on the ancilla qubit state~$\rho_\textsc{q}$ has the Kraus form \begin{equation}
		\mathcal{N}_{\mathrm{damp}}[\rho_\textsc{q}]=\sum_{j=0}^1K_j\rho_\textsc{q}\,K_j^\dagger.
	\end{equation}
	The no-jump operator \mbox{$K_0=\ket{0}\!\!\bra{0}+\sqrt{1-p}\ket{1}\!\!\bra{1}$} is the same for both channels. The jump operator is \mbox{$K_1^{(\mathrm{amp})}=\sqrt{p}\ket{0}\!\!\bra{1}$} for amplitude damping and \mbox{$K_1^{(\mathrm{ph})}=\sqrt{p}\ket{1}\!\!\bra{1}$} for phase damping. We consider amplitude and phase damping of equal strength~$p$ on the ancilla throughout this~article. 
	
	Anticipating the qutrit extension in Sec.~\ref{sec:qutrit}, we also model the noise on a qutrit ancilla with three eigenstates $\ket{j}$ where $j\in\{0,1,2\}$, corresponding to the lower, middle, and upper energy levels of a three-level system respectively. For this system, the cascaded amplitude-damping channel with a single parameter $p$ can be described simply by the Kraus operators
	\begin{align}
		K_0=&\,\ketbra{0}{0}+\sqrt{1-p}\ketbra{1}{1} + \sqrt{1-p}\ketbra{2}{2},\notag\\
		K_1=&\,\sqrt{p}\ketbra{1}{2}, K_2=\sqrt{p}\ketbra{0}{1},
		\label{eq:qutrit-AD}
	\end{align}
	whereas a specialized phase-damping on the $\ketbra{1}{1}$ state can be described by
	\begin{align}
		K_0=&\,\ketbra{0}{0}+\sqrt{1-p}\ketbra{1}{1} +\ketbra{2}{2},\notag\\
		K_1=&\,\sqrt{p}\ketbra{1}{1}.
		\label{eq:qutrit-PD}
	\end{align}
	The above-mentioned Kraus operators do not commute, but a short calculation \mbox{verifies that the two channels commute.}

	\section{Single qubit-assisted code-agnostic noise suppression}
	\label{sec:single_ancilla}
	
	The hybrid interferometer we consider here (Fig.~\ref{fig:protocols}(a)) is represented by the output state $\rho_{\textsc{b},\mathrm{supp}}\propto\, \bra{0}\Us^\dag\,\mathcal{N}[\Us\,\rho_\textsc{b}\otimes\ket{0}\!\!\bra{0}\Us^\dag]\,\Us\ket{0}$, where $U_\mathrm{s}$ is the suppression unitary, $\mathcal{N}$ is the noise affecting the hybrid CV-DV state, and the normalization gives the success~probability of the protocol. 
	
	Writing the channel $\mathcal{N}$ in Kraus form, consider a pair of operators corresponding to $l$-photon losses
	\mbox{$A_l:=\sqrt{\frac{\mu^l}{l!}}\sqrt{1-\mu}^{\AdA}a^l$}
	and $k$-excitations
	\mbox{$B_k:=\sqrt{\frac{(1-G^{-1})^k}{k!\, G}}\,a^{\dagger k}G^{-\frac{\AdA}{2}}$}~\cite{Ivan2011:Operator-sum, Gagatsos2017:Bounding}.
	Under the action of the suppression unitary
	$\Us=\E{\I\vartheta\,a^\dag a \,\widehat{\mathbf{n}}\cdot\rvec{\sigma}}$, where $\rvec{\sigma}=(\sigma_x,\sigma_y,\sigma_z)^\top$ is the column of Pauli operators and $\nhat$ is a unit-vector,
	and its adjoint, the paired Kraus operators transform into
	\begin{align}
		\Us^\dagger B_k A_{l} \Us=&c_{l,k}\sqrt{\frac{1-\mu}{G}}^{\AdA}a^{\dagger k}a^l\E{\I\vartheta(l-k)\nhat\bm{\cdot}\bm{\sigma}},
		\label{eq:suppressed_Kraus}
	\end{align}
	where $c_{l,k}:=\sqrt{\left(\frac{G-1}{1-\mu}\right)^k\frac{\mu^l}{k!\,l!\,G}}$. For convenience, we define three new variables as $x:=(1-\mu)/G$,  $y:=\mu G/(1-\mu)$ and $z:=1-1/G$.
	
	The ancilla is initialized and finally, conditionally measured in the same state~$\ket{0}\!\!\bra{0}=(1+\khat\bm{\cdot}\bm{\sigma})/2$ with $\khat$ the Bloch-sphere unit vector along $+z$, unless stated otherwise. For a two-level system like an atom, this is the ground state and a natural choice for the ancilla qubit, as it remains stable \mbox{under~damping~noise}. Such a conditional measurement results in an unnormalized output
	\begin{align}
		\tilde{\rho}_\textsc{b}'=\sum_{k,l\geq0}&L_{l,k}\,\rhob\,L_{l,k}^\dagger
		|\opinner{0}{\E{\I\vartheta (l-k)\AdA\, \nhat\bm{\cdot}\bm{\sigma}}}{0}|^2.
		\label{eq:supp-state}
	\end{align}
	For notational convenience we define \mbox{$L_{l,k}:=c_{l,k}\sqrt{x}^{\AdA}a^{\dagger k}a^l$}.
	Equivalently, $U_\mathrm{s}$ can be written as two single-qubit Hadamard gates sandwiching a conditional $Z$-rotation~\cite{Nielsen2010:Quantum}.
	
	For any $\rho_\textsc{b}$, the configuration corresponding to the CF gate,~$\vartheta=\pi/2$ and $\nhat\perp\khat$, completely cancels all paired photon-loss and -gain events that alter the photon-number parity, assuming a perfect DV ancilla. The name controlled-Fourier is used here as notational convenience to distinguish this $\vartheta=\pi/2$ choice from the more general controlled-rotation primitive of~\cite{Heeres2015:Cavity, Krastanov2015:Universal}. 
	
	To verify this, the DV inner product evaluates to
	\begin{align}
		&\,\opinner{0}{\E{\I\frac{\pi}{2} (l-k) \AdA\, \nhat\bm{\cdot}\bm{\sigma}}}{0}\notag\\
		&=\cos((l-k)\pi/2)+\I \opinner{0}{\nhat\bm{\cdot} \bm{\sigma}}{0} \sin((l-k)\pi/2)\notag\\
		&=\cos((l-k)\pi/2)\notag\\
		&\therefore|\opinner{0}{\E{\I\frac{\pi}{2} (l-k) \AdA\, \nhat\bm{\cdot}\bm{\sigma}}}{0}|^2=\delta_{(l-k)\bmod 2}.
		\label{eq:odd-suppressed}
	\end{align}
	This suppresses noise events that lead to a change in the photon-number parity. This mechanism differs from standard error correction, where parity measurement occurs entirely after the action of the noise channel without a prior entangling operation.
	
	A miscalibration $\vartheta=\pi/2+\epsilon$ causes imperfect suppression. The DV factor $\cos^2((l-k)\vartheta)$ for odd $(l-k)$ leaks quadratically with $\epsilon$ as $\sin^2((l-k)\epsilon)\cong(l-k)^2\epsilon^2$ rather than vanishing. For the rest of the article we consider perfect calibration and focus on the noise between the suppression unitaries.
	
	The dispersive interaction required to implement $\Us$ is a well-established, high-fidelity operation used for parity measurements in circuit QED~\cite{Sun2014:Tracking}. Our protocol arranges these gates in a sandwich geometry across a potentially transmitting channel as in~\cite{Reiserer2014:Quantum, Kurpiers2018:Deterministic, Gonzales2025:Detecting}. This shifts the focus from post-noise parity measurement to parity filtration \emph{across} a channel~\cite{Gisin2005:Error}, allowing potentially distributed error suppression independent of the input. The DV ancilla also serves as a non-Gaussian resource. No Gaussian, linear-optical operation can act on photon-number parity.
	
	Such cancellation of paired losses and gains leads to noise suppression for \emph{any} input state~$\rho_\mathrm{in}$. By averaging over Haar-random~\cite{Harrow2013:Church, Mele2024:Introduction} pure qubit input states encoded on the \mbox{orthonormal} bosonic logical kets as $\ket{\psi_\mathrm{in}}=c_0\ket{0_\textsc{l}}+c_1\ket{1_\textsc{l}}$, the \emph{average fidelity} $\overline{\mathcal{F}}=\AVG{c_0,c_1}{\opinner{\psi_{\mathrm{in}}}{\rho_\mathrm{out}}{\psi_{\mathrm{in}}}}$, over real~$c_0$ and complex~$c_1$ on the surface of the qubit Bloch~sphere, reflects the average quality of encoded output physical states. The cancellation of paired losses and gains manifests as the absence of linear order terms in this average fidelity. Physical-level fidelity improvements translate to logical fidelity gains under an outer code, beyond the scope of this article.

	We present concrete examples from common single-mode bosonic codes~\cite{Albert2018:Performance, Grimsmo2020:Quantum, Joshi2021:Quantum, Teo2025:Linear}. The cat codes $\texttt{cat}(n,\alpha)$ are superpositions of $n$ coherent states of amplitude $\alpha$ on a ring~\cite{Cochrane1999:Macroscopically, Jeong2002:Efficient, Ralph2003:Quantum, Leghtas2013:Hardware-Efficient, Shringarpure2024:Error}. The binomial codes $\texttt{bin}(n,\kappa)$ are superpositions of maximum $\kappa$, $n$-gapped Fock states distributed binomially~\cite{Michael2016:New}. The finite-energy approximate Gottesman--Kitaev--Preskill codes $\texttt{gkp}(\Delta)$ are superpositions of displaced-squeezed states with a damping factor $\Delta$~\cite{Gottesman2001:Encoding}. As our scheme is assisted with a DV ancilla, we plot all performance metrics as a function of the DV ancilla noise strength throughout this article.

	\section{Transformation to Fock-diagonal noise with multiple ancilla}
	\label{sec:multi_ancilla}
	
	When the procedure is supplemented with multiple ancillas, thermal and Gaussian displacement noise channels are transformed into a mixture of Fock-diagonal channels. The state in Eq.~\eqref{eq:odd-suppressed} is generalized easily to an arbitrary number of ancillas $K$ with the phase angles $\vartheta_j$ with $j=1,2,\ldots, K$. The general output state is then given by
	\begin{align}
		\tilde{\rho}_{\textsc{b}}^\prime=\sum_{k,l\geq0}&L_{l,k}\,\rhob\,L_{l,k}^\dagger\prod_{j=1}^{K} |\opinner{0}{\E{\I\vartheta_j (l-k)\AdA\, \nhat\bm{\cdot}\bm{\sigma}}}{0}|^2.
		\label{eq:multiancilla_general_theta}
	\end{align}
	
	With the choice of angles $\vartheta_j:=\pi/2^j$ and $\nhat\perp\khat$, the product simply reduces to the Kronecker delta $\delta_{(l-k)\bmod2^K}$ giving the unnormalized output state
	\begin{equation}
		\tilde{\rho}_\textsc{b}^\prime=\sum_{k,l\geq0}L_{l,k}\,\rhob\,L_{l,k}^\dagger\delta_{(l-k)\bmod 2^K},
		\label{eq:multianc-output}
	\end{equation}
	whose trace gives the success probability
	\begin{align}
		&p_{\rm{succ}}=\notag\\
		&\mathop{\mathrm{tr}}\bigg\{\rhob\bigg[x^{\AdA}\frac{1-z}{2^K}\sum_{j=0}^{2^K-1}
		\frac{\bigg(1+y\omega^j+\frac{z\omega^{- j}}{1-z\omega^{-j}}\bigg)^{\AdA}}{1-z\omega^{-j}}\bigg]\bigg\},
		\label{eq:psucc_K}
	\end{align}
	where $\omega:=\E{\I {\pi}/{2^{K-1}}}$.
	As it is linear in the density operator, the average success probability is obtained by simply replacing $\rhob$ with $\SL$, which is the normalized identity of the bosonic codespace (see Appendix~\ref{app:Haar_average}).		
	
	\begin{figure}[htbp!]
		\centering
		\includegraphics[width=0.9\columnwidth]{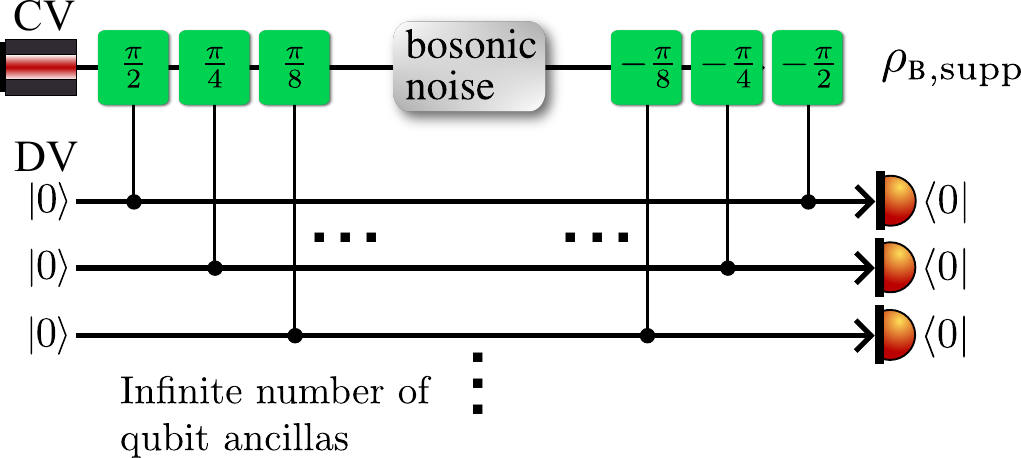}
		\caption{\label{fig:multiqubit_extension} Multiqubit extension for converting thermal or Gaussian displacement noise, assumed to be solely between the suppression unitaries, into a mixture of Fock-diagonal noise. For pure photon loss, this setup furnishes a coherent Fock damping on the input so that the output is an attenuated pure state.
		}
	\end{figure}
	
	In the asymptotic limit ($K\to\infty$), the success probability converges to a non-zero constant (see Appendix~\ref{app:psucc_singlemodesupp})
	\begin{equation}
		p_{\rm{succ}}=\frac{1}{G} \sum_{n\geq0} \bra n \rho_{\textsc{b}} \ket n \bigg(\frac{1}{G}-\mu\bigg)^{n}P_{n}\bigg(\frac{1-\mu(2-G)}{1-\mu G}\bigg),
	\end{equation}
	where $P_n(\cdot)$ represents the $n$th Legendre polynomial. This non-vanishing probability parallels findings of asymptotically large linear optical setups~\cite{Teo2025:Linear}. The factor of $\delta_{(l-k)\bmod 2^K}$ in Eq.~\eqref{eq:multianc-output} becomes $\delta_{l-k}$, leaving only the photon-number-conserving Kraus operators $L_{l,l}$ and a \mbox{\emph{Fock-diagonal}} residual noise, acting as a filter.

\begin{figure}[htbp!]
		\centering
		\includegraphics[width=\columnwidth]{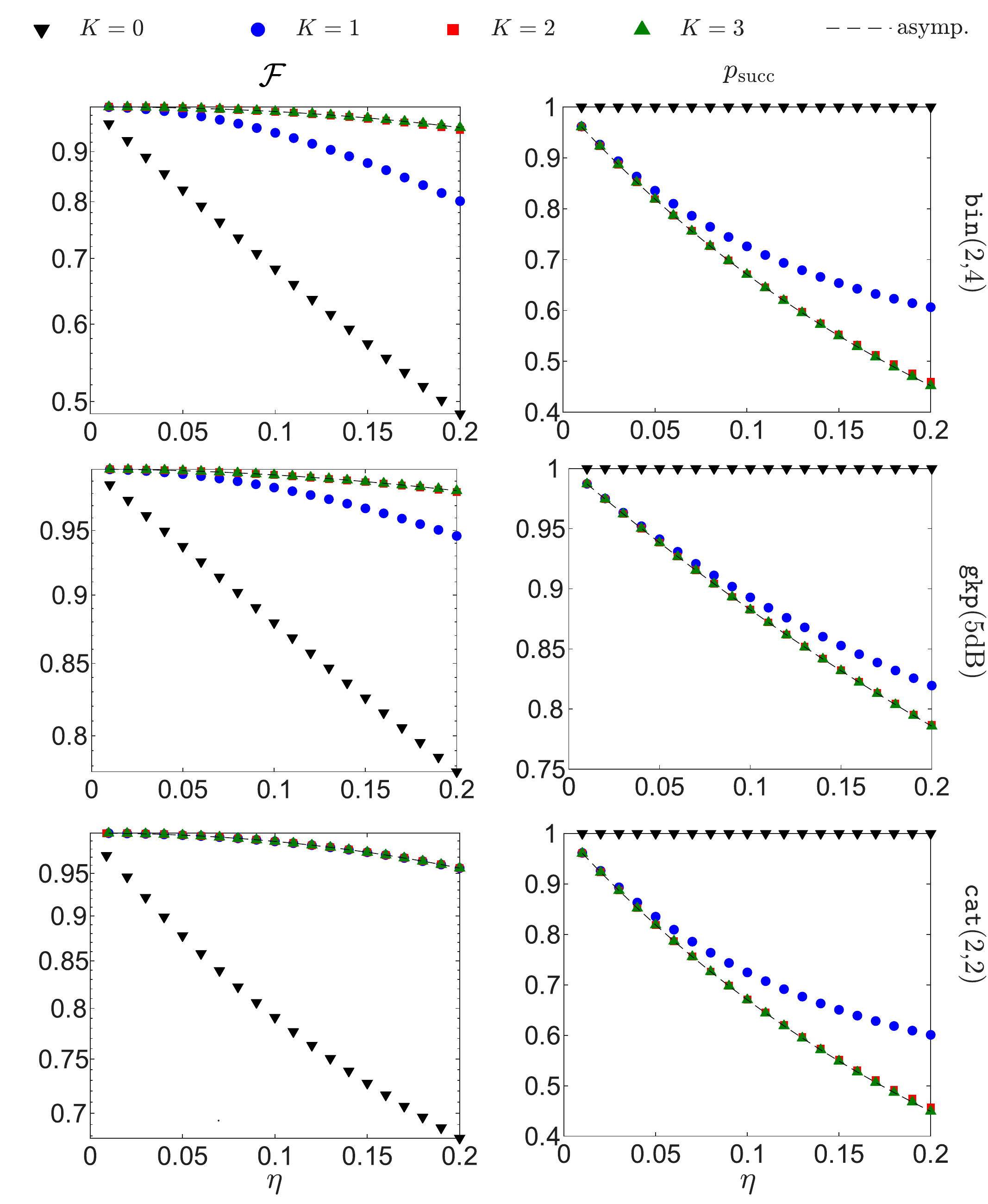}
		\caption{\label{fig:multiqubit_eta} Average suppression performances as a function of the photon loss rate $\eta$ for varying numbers of DV ancillas $K$. The solid lines ($K=1, 2, 3$) demonstrate rapid convergence toward the theoretical asymptotic limit. Perfect ancillas have been assumed in order to focus on the effects of adding the ancillas alone here.}
	\end{figure}

	For experimental feasibility, the convergence to this limit is extremely rapid. Just $K=2$ or $K=3$ ancillas achieve suppression fidelities and success rates practically indistinguishable from the asymptotic limit for typical bosonic codes, as shown in Fig.~\ref{fig:multiqubit_eta}.
	We rigorously bound the gap between the $K$-ancilla protocol and the asymptotic $K \to \infty$ limit, assuming perfect gates for the analysis.

	The difference in the $K$-ancilla success probability from the asymptotic limit is the sum of terms with $l \neq k$ but $k = l \bmod M$, where $M := 2^K$, so that
	\begin{align}
		\Delta p(K) &= p_{\mathrm{succ}}(K) - p_{\mathrm{succ}}(\infty) \notag\\
		&= \mathop{\mathrm{tr}}\bigg\{\frac{\rho_{\textsc{b}}}{G}x^{\AdA}\sum_{\substack{l\neq k\geq0 \\ k\equiv {l}\bmod {M}}}\frac{y^l}{l!}a^{\dagger l}\left[\frac{z^k}{k!}\,a^ka^{\dagger k}\right]a^l\bigg\}.
	\end{align}
	The dominant terms correspond to $(l=M, k=0)$ and $(l=0, k=M)$.
	
	For photon loss ($z=0$ or $G=1$), $k=0$ and therefore the protocol only fails to reject losses in multiples of $M$, and so the error is dominated by the $l = M = 2^K$ term, which gives
	\begin{equation}
		\Delta p(K) \cong \frac{y^{2^K}}{(2^K)!} \mathop{\mathrm{tr}}\{\rho_{\textsc{b}}\, x^{\AdA}:(\AdA)^{2^K}:\}.
	\end{equation}
	Since $a^{\dagger l} a^l = 0$ for any Fock state $\ket{n}$ with $n < l$, any bosonic code with a maximum photon number $N_{\max}$ (such as $\texttt{bin}(n, \kappa)$) has $\Delta p(K) = 0$ exactly whenever $K = \lceil \log_2(N_{\max}+1) \rceil$, since losses of $2^K$ or more photons are then physically impossible.
	
	Under thermal noise, $k\geq0$ is unbounded but evaluating the sum of dominant terms ($l=2^K, k=0$) and ($l=0, k=2^K$), upto the leading-order we have
	\begin{align}
		&\Delta p(K) \notag\\
		&\cong \frac{1}{(2^K)!} \mathop{\mathrm{tr}}\bigg\{ \frac{\rho_\textsc{b}}{G} x^{\AdA} \left[ y^{2^K}:(\AdA)^{2^K}:+z^{2^K} \vdots(\AdA)^{2^K}\vdots \right] \bigg\}.
	\end{align}
	For small loss parameter $y \cong \mu$ and sufficiently small thermal gain parameter $z \cong \eta\bar{n}$, the deviation $\Delta p(K)$ decays rapidly as $\mathcal{O}( (\eta(1+\bar{n}))^{2^K} / (2^K)! )$, making the success probability practically indistinguishable from the nonzero, asymptotic limit.

	The same machinery bounds the convergence in the unnormalized fidelity $\widetilde{\mathcal{F}}(K):=\opinner{\psi_\mathrm{in}}{\tilde\rho'_\textsc{b}}{\psi_\mathrm{in}}$, for a unit-norm pure code input $\ket{\psi_\mathrm{in}}$. For any positive operator $X\geq 0$, $\opinner{\psi_\mathrm{in}}{X}{\psi_\mathrm{in}}\leq\tr{X}$ and each term with $l\neq k$ in the fidelity, $\opinner{\psi_\mathrm{in}}{L_{l,k}\rho_\textsc{b}L_{l,k}^\dagger}{\psi_\mathrm{in}}$, is positive, hence $\Delta\widetilde{\mathcal{F}}(K)\,\leq\,\Delta p(K)$.
	The normalized post-selected fidelity $\mathcal{F}(K):=\widetilde{\mathcal{F}}(K)/p_\mathrm{succ}(K)$ with $p_\mathrm{succ}(K)> p_\mathrm{succ}(\infty)$ satisfies
	\begin{equation}
		\Delta\mathcal{F}(K)\,=\,\frac{\widetilde{\mathcal{F}}(K)}{p_\mathrm{succ}(K)}\,-\,\frac{\widetilde{\mathcal{F}}(\infty)}{p_\mathrm{succ}(\infty)}\,<\,\frac{\Delta\widetilde{\mathcal{F}}(K)}{p_\mathrm{succ}(\infty)},
	\end{equation}
	and inherits the same rate of convergence as $\Delta p(K)$.
	
	\section{Quadratic scaling of the noise-suppressed fidelity}
	\label{sec:linear_order_suppression}
	
	Here we show that for the CF-based suppression protocol with a perfect DV ancilla, the fidelity after suppression does not contain linear order terms. We drop the subscript on the qumode density operator for convenience. We begin with the coefficients in Eq.~\eqref{eq:supp-state}, expanded up to the second order of the Taylor series in the two parameters as
	\begin{align}
		&c^2_{0,0}\cong1-z, &&c^2_{0,2}\cong\frac{\mu^2}{2},&& c^2_{2,0}\cong\frac{z^2}{2},\notag\\
		&c^2_{1,1}\cong z\mu,
		&&c^2_{1,0}\cong(1+z)\mu,&&c^2_{0,1}\cong z(1+\mu),
	\end{align}
	and the \emph{rescaling} operator
	\begin{align}
		&\sqrt{x}^{\AdA}\cong1-(\mu+z) T_1+\mu z T_{2,1}+(z^2+\mu^2) T_{2,2},
	\end{align}
	with the definitions \mbox{$T_1:={\AdA}/{2}$}, \mbox{$T_{2,1}:={(\AdA)^2}/{4}$}, and \mbox{$T_{2,2}:=\AdA(\AdA-2)/8$}.
	
	Using these approximations while restricting to a maximum of two noisy jumps, together with the observation from Eq.~(\ref{eq:odd-suppressed}), for a noiseless DV ancilla,
	we obtain the normalized state 
	\begin{align}
		\rho'&\cong\rho+z\big(2\MEAN{T_1}\rho-\{T_1,\rho\}\big)\notag\\
		&+\mu\big(2\MEAN{T_1}\rho-\{T_1,\rho\}\big)+\mu z R_{1,1}+z^2\,R_{2,0}+\mu^2\,R_{0,2},
	\end{align}
	where the quantities $R_{j,k}$ are independent of the noise parameters~(see Appendix~\ref{app:linear_order_suppression}).
	The fidelity is given by $\tr{\rho \rho'}$ for a pure input state $\rho$. Thus linear terms in $\mu$ and $z$ vanish on using $\rho^2=\rho$ and the cyclic property of the trace. This result applies to both thermal noise and Gaussian displacement noise with appropriate choices of $\mu$ and $z\,(=1-1/G)$ (see Appendix~\ref{app:noise-sources}). A numerical demonstration of the resulting $\eta^2$ scaling, together with a head-to-head comparison against active input squeezing, is deferred to Sec.~\ref{sec:comparison} (Fig.~\ref{fig:mu_z}).

	Specializing to thermal noise of error rate $\eta$, the channel is realized as a quantum-limited amplifier of gain $G:=1+\eta\,\bar{n}$ following a photon-loss channel with rate $\mu:=1-(1-\eta)/G$ (see Appendix~\ref{app:noise-sources})~\cite{Ivan2011:Operator-sum,Gagatsos2017:Bounding}.
	Following a similar procedure and combining with identities for averages over the Haar measure of unitaries (see Appendix~\ref{app:Haar_average}), we obtain
	\begin{align}
		&\,\overline{\mathcal{F}}_\mathrm{supp}\cong1-\eta^2\bigg\{\bar{n}^2+3\,\left(\bar{n}+\frac{1}{2}\right)^2\,\tr{\SL\,(a^\dag a)^2}\notag\\
		&\,+\left(\bar{n}^2-\bar{n}-\frac{1}{2}\right)\tr{\SL\,a^\dag a}-\left[\frac{1}{6}+\frac{4}{3}(\bar{n}^2+\bar{n})\right]g(a^\dag a)\notag\\
		&\,-\left[\frac{1}{3}+\frac{2}{3}\left(\bar{n}^2+\bar{n}\right)\right]g(a^2)\bigg\},
		\label{eq:Fsupp}
	\end{align}
	where $g(Y):=\tr{\SL\,Y\SL\,Y^\dag}+|\tr{\SL\,Y}\!|^2$ and $\SL:=\big(\ket{0_\textsc{l}}\!\!\bra{0_\textsc{l}}+\ket{1_\textsc{l}}\!\!\bra{1_\textsc{l}}\big)/2$ is the normalized codespace identity defined for convenience, $\eta$ is the thermal loss rate and $\bar{n}$ is the mean number of thermal photons.

	The suppression protocol results in an error-rate scaling ($\eta^2$) of the infidelity~$1-\overline{\mathcal{F}}_\mathrm{supp}$.
	This offers an advantage over the unsuppressed case where the corresponding infidelity scales linearly with $\eta$~(see Appendix~\ref{app:psucc_CVcomm}).
	
	The corresponding \emph{average success probability} reads (see Appendix~\ref{app:psucc_singlemodesupp})
	\begin{equation}
		\overline{p}_\mathrm{succ}=\frac{1}{2}+\frac{1}{2(2G-1)}\tr{\SL
			\left(\frac{1-2\mu G}{2G-1}\right)^{\AdA}}.
		\label{eq:psucc}
	\end{equation}
	As $\SL\geq0$ and $G\geq1$, $\overline{p}_\mathrm{succ}\geq0.5$ when $\mu G\leq0.5$.

	\section{Protection of hybrid entangled states}
	\label{sec:hybrid_entangled}
	
	The simple single-ancilla assisted protocol also protects the bosonic mode of hybrid-entangled states~\cite{Bose2024:Long-distance} (such as $\ket{\alpha}\!\!\ket{0}+\ket{\beta}\!\!\ket{1}$) with just one additional qubit ancilla. 
	
	Consider hybrid states of the form 
	\mbox{$
		\ket{\psi}_{\textsc{bq}}=\left({\ket{\alpha}\!\!\ket{0}\pm\ket{\beta}\!\!\ket{1}}\right)/{\sqrt{2}},
		$}
	which have the density operator
	\begin{align}
		\rho_{\textsc{bq}}=&\,\frac{1}{2}\big(\ket{\alpha}\!\!\bra{\alpha}\otimes\ketbra{0}{0}+\ketbra{\beta}{\beta}\otimes\ketbra{1}{1}\notag\\
		&\,\pm\ketbra{\alpha}{\beta}\otimes\ketbra{0}{1}\pm\ketbra{\beta}{\alpha}\otimes\ketbra{1}{0}\big).
	\end{align}
	where $\ket{\alpha}$ and $\ket{\beta}$ are two distinct coherent states.
	
	As an example, consider photon losses where we have for the $l$th-loss Kraus operator sandwiched by the suppression unitaries as
	$
	\Us^\dagger \,K_l\,\Us=\sqrt{\frac{\eta^l}{l!}}\,(1-\eta)^\frac{\AdA}{2}\,a^l\,\E{\I\, l \vartheta\,\nhat\bm{\cdot}\bm{\sigma}},
	$
	where the conditional rotation acts between the qumode of the hybrid state and the qubit ancilla for the suppression protocol.
	 Then, the arbitrary dyadic components of the density operator are modified as
	\begin{align}
		&\ketbra{\alpha}{\beta}\otimes\ketbra{\psi}{\phi}\otimes\ketbra{0}{0}_{\text{anc}}\notag\\
		&\mapsto \sum_{l\geq0} \frac{\eta^l}{l!}\ket{\sqrt{1-\eta}\alpha}\ket{\psi}(\alpha\beta^*)^l\E{-\frac{\eta}{2}(|\alpha|^2+|\beta|^2)}\notag\\
		&\qquad\quad\bra{\phi}\!\bra{\sqrt{1-\eta}\beta}\otimes\E{\I\,l\vartheta\,\nhat\bm{\cdot}\bm{\sigma}}\,\ketbra{0}{0}_{\text{anc}}\,\E{-\I\,l\vartheta\,\nhat\bm{\cdot}\bm{\sigma}},
	\end{align}
	where \mbox{$\ketbra{0}{0}_{\text{anc}}=\big(1+\khat\bm{\cdot}\bm{\sigma}\big)/{2}$} is the initial ancilla state.
	Projecting only the ancilla in the same state gives the familiar $|\opinner{0}{\E{\I\,l\vartheta\,\AdA}}{0}|^2_{\text{anc}}=\delta_{\text{even}(l)}$ for $\vartheta=\pi/2$ and $\nhat \perp \khat$, which again cancels the linear terms in $\eta$ in the output state. The same argument extends to thermal and Gaussian displacement noise. It also covers hybrid dyads $\ketbra{\alpha}{\beta}\otimes\ketbra{\psi}{\phi}$ between a qumode and an arbitrary system.
	
	\section{Conversion to error-detection protocol for rotation-symmetric codes}
	\label{sec:RSBC}

	The protocol applies to any single-mode bosonic input state. Codes with explicit parity structure admit a simpler analysis with additional resilience to ancilla damping noise. Consider families of bosonic codes stabilized by the parity operator,
	\mbox{$[\ket{0_\textsc{l}}\!\!\bra{0_\textsc{l}},(-1)^{a^\dag a}]=0=[\ket{1_\textsc{l}}\!\!\bra{1_\textsc{l}},(-1)^{a^\dag a}]$},
	with logical states having either like or opposite photon-number parities.
	
	A code is like-parity when both logical codewords share the same photon-number parity eigenvalue $\pm 1$ under $(-1)^{\AdA}$, and opposite-parity when the eigenvalues differ. The first CF~unitary gate reads $ U_\mathrm{s}=\cos(\pi\AdA/2)+\I\sin(\pi\AdA/2)\nhat\bm{\cdot}\bm{\sigma}$,
	and only the first term survives for like (even)-parity codes as $\cos(\pi\AdA/2)=\sum_{k\geq0}(-1)^k\ket{2k}\!\!\bra{2k}$ lives in the even subspace. This implies that the ancilla remains in the ground state, unaffected. It remains stable as such under damping noise up to the action of the second unitary $U_\mathrm{s}^\dag$, 
	which subsequently flips the DV state \emph{only} if the bosonic noise alters photon-number~parity.
	\begin{figure}
		\centering
		\includegraphics[width=\columnwidth]{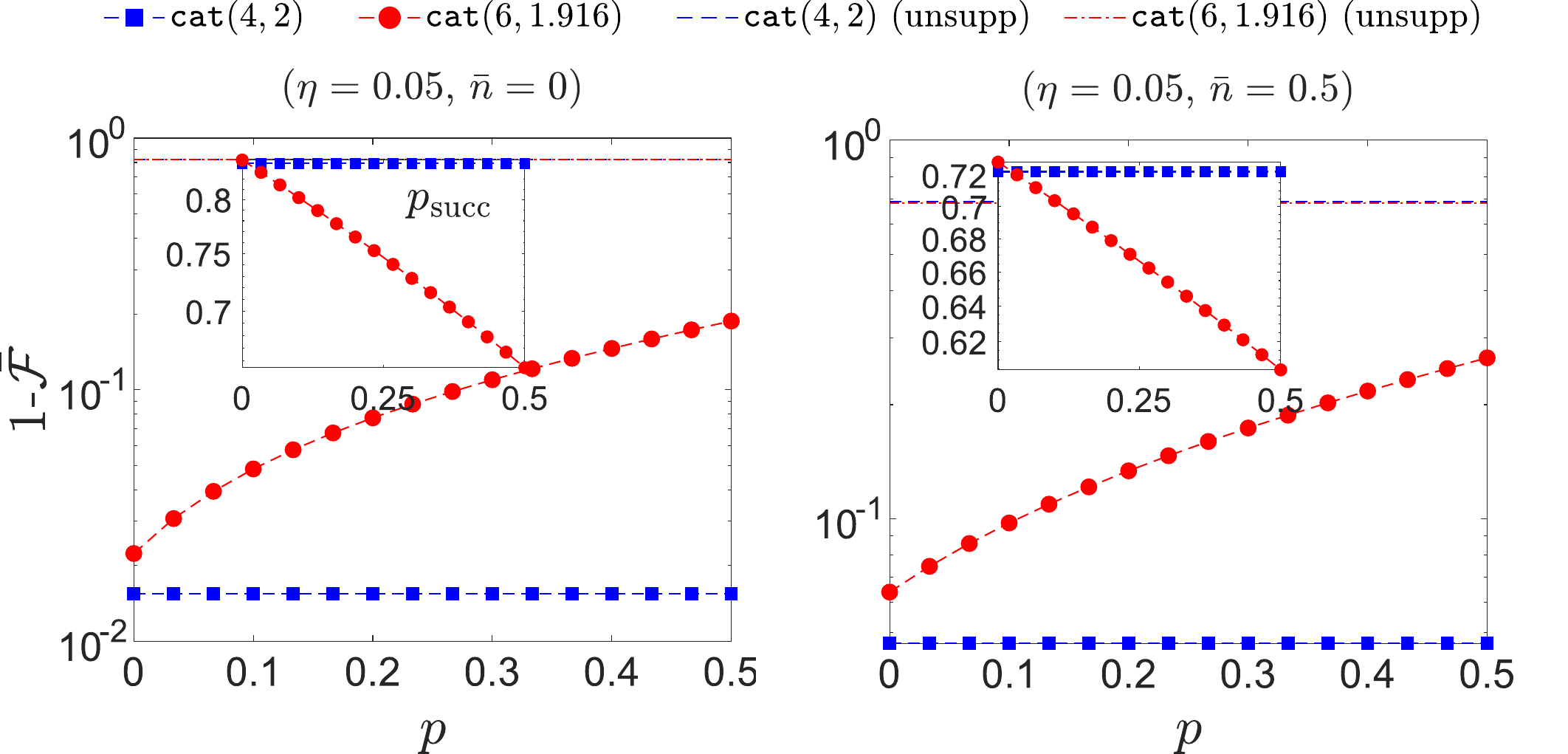}
		\caption{\label{fig:likevsopp}
			Comparison of average suppression performance with conditional Fourier~(CF) gates between a like (even)-parity code $\texttt{bin}(2,4)$ and an opposite-parity bosonic code: $\texttt{cat}(6,1.916)$, both of which have similar average Gaussian moments of the photon-number distribution~$(\langle \AdA\rangle\cong 4$, \mbox{$\langle (\AdA)^2 \rangle\cong 20$}, \mbox{$\langle a^2\rangle=0)$} with respect to the state~$C_\textsc{l}$. 
			The former is more resilient to the composite amplitude and phase damping qubit noise of equal strengths~$p$. 
			The dashed and dot-dashed lines represent their respective, roughly \mbox{identical~performances} without suppression. These are flat because no ancillas were used. Insets show average success~probability. }
	\end{figure}
	\begin{figure}[htbp!]
		\centering
		\includegraphics[width=\columnwidth]{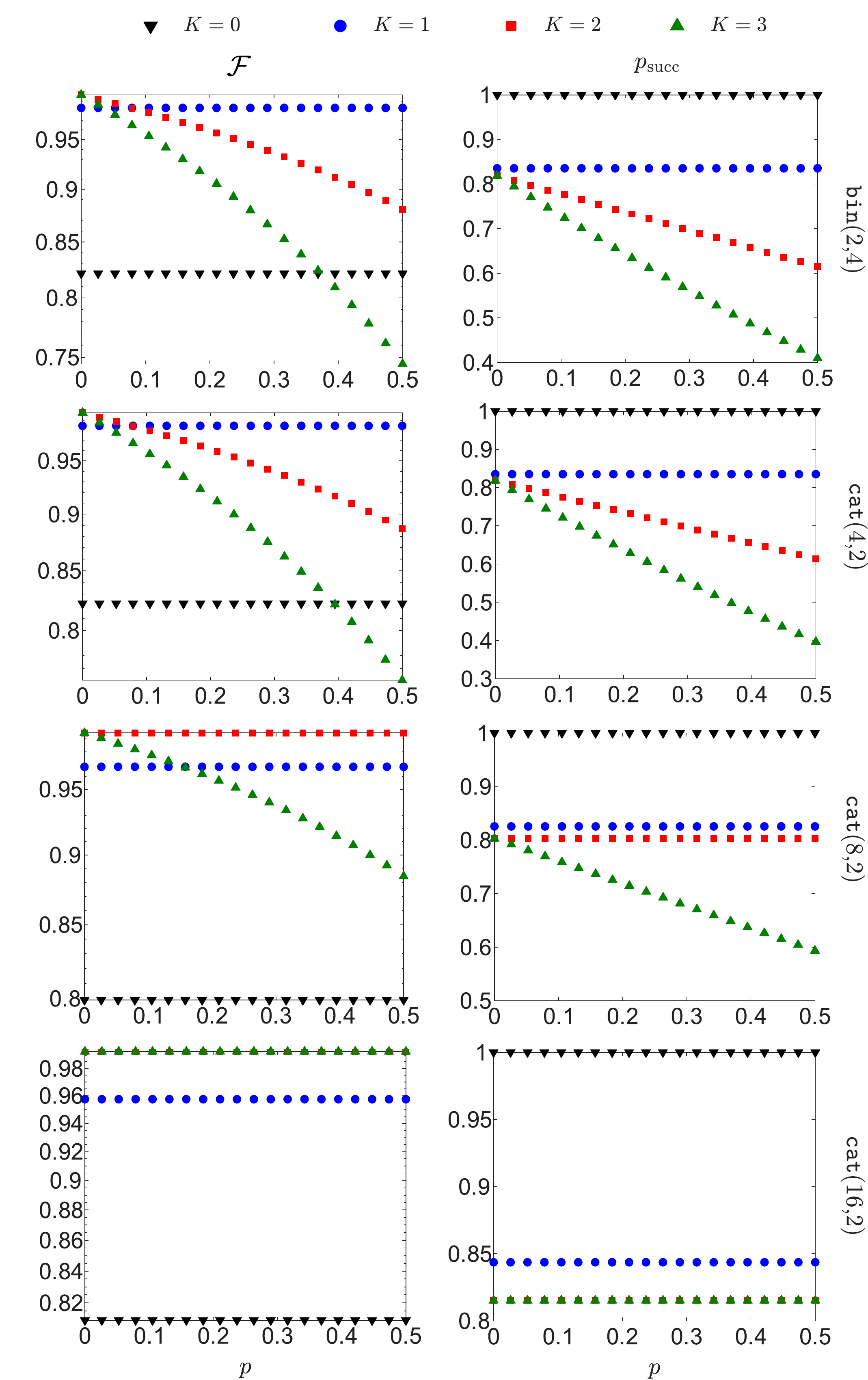}
		\caption{\label{fig:multiqubit_p} Average suppression performances as a function of the composite DV damping noise strength $p$ for varying numbers of ancillas $K$. The multi-ancilla protocol naturally reduces to pure error detection for rotation-symmetric bosonic codes supported on Fock states that are multiples of $2^K$. The conventional $\texttt{cat}(2^{K+1},\alpha)$ codes furnish such an example. This simplification is responsible for the complete resilience to uncalibrated DV damping noise, manifested as the perfectly flat fidelity curves. A fixed photon loss rate of $\eta=0.05$ is assumed.}
	\end{figure}
	
	As the second term plays no role, we can further simplify the setup by substituting the first conditional rotation with a single, local, bosonic rotation by $\pi/2$ for an identical performance as shown in Fig.~\ref{fig:likevsopp}. A similar argument applies to like (odd)-parity codes when the DV ancilla is initialized in the state $\nhat\bm{\cdot}\bm{\sigma}\ket{0}$ instead. In this case, the ancilla state after the first unitary is~$\ket{0}$ again, leading to the same robustness under DV damping~noise. These are the rotation-symmetric bosonic codes~\cite{Grimsmo2020:Quantum} with at least 2-fold rotation symmetry. 
	
	Generalizing this observation to $2^K$-fold rotation symmetric codes, the appropriate angles for complete resilience to DV damping noise with $K$-ancillas of Fig.~\ref{fig:multiqubit_extension} follow $\vartheta_j=\pi/2^j$ with $j\in\{1,2,\cdots,K\}$ as shown by the results in Fig.~\ref{fig:multiqubit_p}.	
	For such rotation-symmetric bosonic codes, our scheme reduces to the usual error detection after noise~\cite{Li2023:Memoryless, Li2024:Performance}, a structure recently exploited for fault-tolerant state preparation of the four-component cat code~\cite{Chen2026:Fault}. This reduction is operationally related to the symmetry expansion framework of Ref.~\cite{Endo2025:Quantum}, though our protocol implements the projection coherently during propagation rather than through classical postprocessing. Our numerical results show that even the output fidelities for states with low symmetry have some resilience to the ancilla damping noise compared to \mbox{other protocols}.

	\section{Comparison with related approaches}
	\label{sec:comparison}

	We now compare our CF-based interferometer against three related strategies. These are active squeezing of the encoded state~\cite{LeJeannic2018:Slowing, Brewster2018:Reduced, Schlegel2022:Quantum, Pan2023:Protecting, Rousseau2025:Enhancing}, an extended single-ancilla family that adds conditional displacements to the CF unitary, and the multi-ancilla bypass protocol of Refs.~\cite{Park2022:Slowing, Hastrup2022:Universal} that transfers the qumode information to DV ancillas before the channel acts.

	\subsection{Combining suppression with active squeezing}
	\label{sec:comparison_squeezing}

\begin{figure}[htbp!]
		\centering
		\includegraphics[width=0.90\columnwidth]{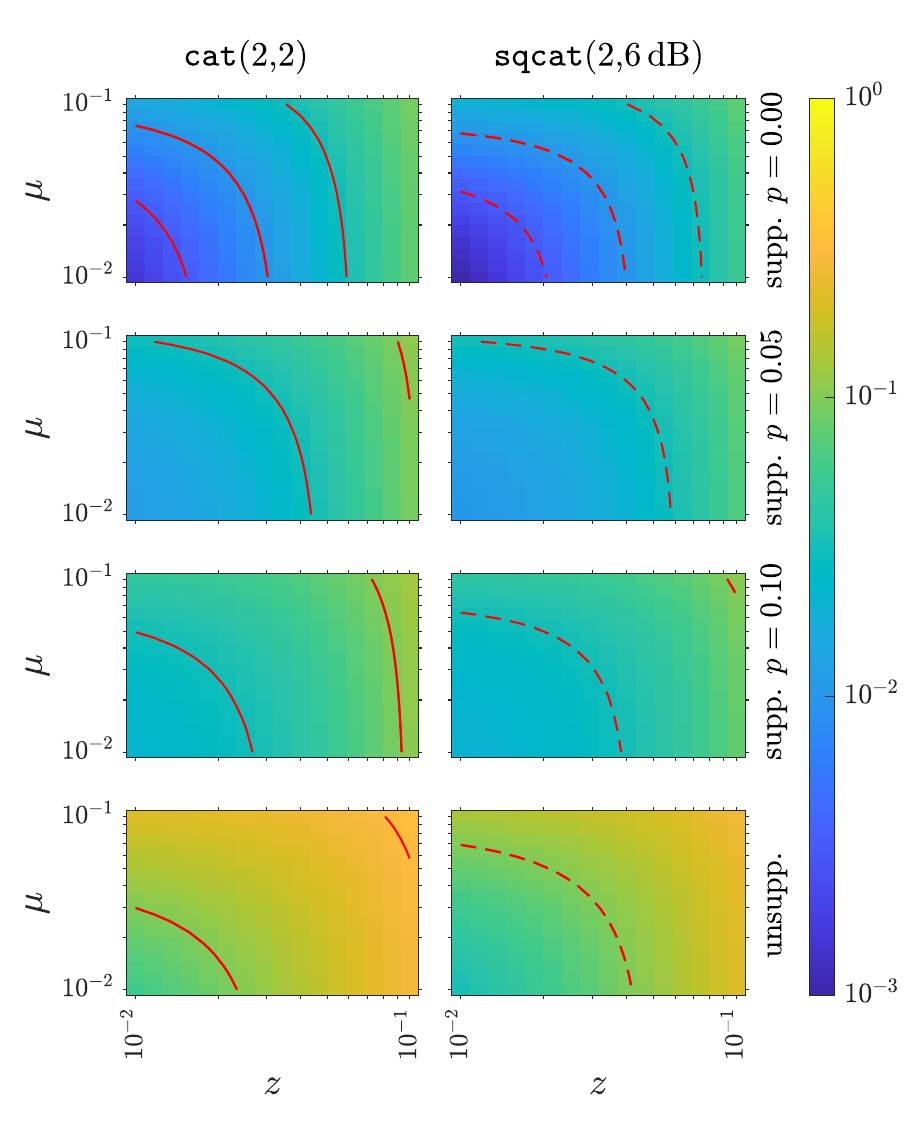}
		\caption{\label{fig:mu_z}
			Average infidelity of the eigenstates of the single-qubit Paulis encoded on the two-component cat code with coherent amplitude of 2 or \texttt{cat}(2,2) and the same with an additional 6dB of squeezing along the real axis applied to the cat codewords, \texttt{sqcat}(2,6dB), with our suppression protocol with a single DV ancilla affected by various damping noise strengths and without suppression. The CV noise is characterized by the loss parameter $\mu$ and the gain parameter $z=1-1/G$. The squeezed cat code retains its advantage over the bare cat code after suppression, demonstrating that active squeezing and the present CF-based protocol can be used in tandem.}
	\end{figure}

	Active squeezing is known to reduce the impact of photon loss on Gaussian and cat-like codewords~\cite{Schlegel2022:Quantum, Pan2023:Protecting}. Acting on the noise channel rather than on the codewords, our suppression protocol can be stacked with active squeezing. To illustrate this stacking, Fig.~\ref{fig:mu_z} compares the two-component cat code $\texttt{cat}(2,2)$ with its squeezed counterpart $\texttt{sqcat}(2,6\,\mathrm{dB})$, in which an additional $6\,\mathrm{dB}$ of squeezing along the real axis is applied to the cat codewords before the channel. For both codes, the noise-suppressed infidelities are well below the unsuppressed reference across a range of CV noise parameters $\mu$ and $z=1-1/G$ and across several strengths of DV ancilla damping $p$. The protocol does not interfere with the gains from squeezing, and squeezing does not undermine the quadratic-in-$\eta$ improvement provided by the \mbox{CF interferometer}.

	The bare cat encoding under our CF protocol achieves lower infidelity than active squeezing alone for DV damping noise up to $p=0.10$ (third-row left versus fourth-row right panels of Fig.~\ref{fig:mu_z}). 

	\subsection{Benchmark with extended single-ancilla schemes}
	\label{sec:comparison_pqp}

	As a benchmark complementary to the single-ancilla framework, we consider a family of suppression unitaries that interleave conditional displacement gates with conditional rotations. The basic building block, sketched in Fig.~\ref{fig:protocols}(b), is a ``PQP-condrot'' layer comprising two conditional displacements in positions~(P) sandwiching a conditional displacement in momenta~(Q), followed by a conditional rotation (condrot). We denote the $L$-fold concatenation of this layer by $[\text{PQP-condrot}]^L$. The CF-only protocol of the present work is the $L=0$ limit. 
	
\begin{figure}
		\centering
		\includegraphics[width=\columnwidth]{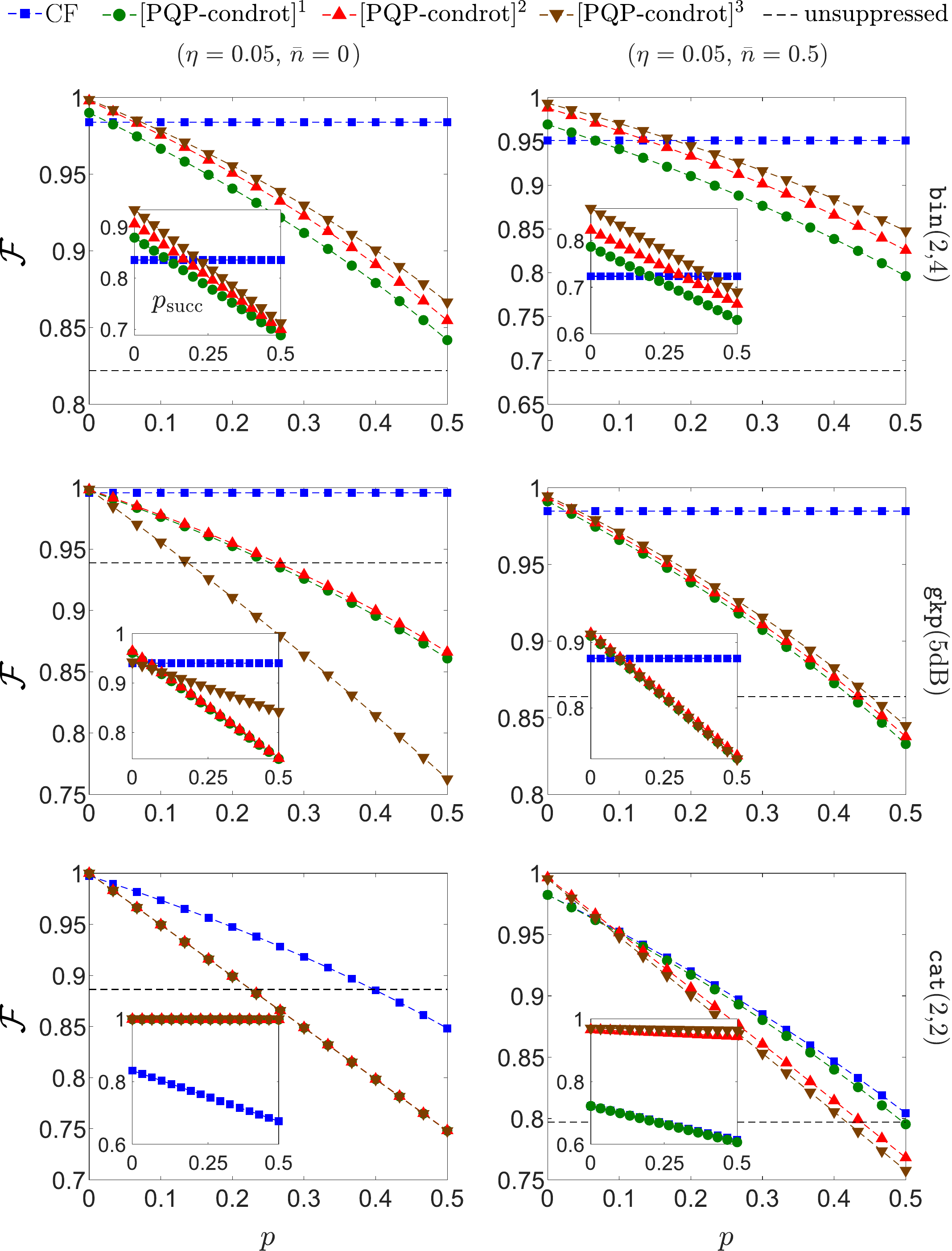}
		\caption{\label{fig:PQPcondrot_damp}
			The CF-gate-only interferometer requires no information about the noise parameters. Its average suppression performance remains impervious to uncalibrated ancilla damping, contrary to a series of conditional displacement gates and conditional rotation gates~([PQP-condrot]$^L$) [see Fig.~\ref{fig:protocols}~(b)], numerically optimized for known parameters of the photon-loss~($\eta=0.05$) and thermal-noise~($\eta=0.05$, $\bar{n}=0.5$) channels. Insets show average success~probability.}
	\end{figure}

	Numerical analysis finds that additional conditional displacements enhance the fidelity if the DV~ancilla is nearly perfect, with some resilience to small DV noise (see Fig.~\ref{fig:PQPcondrot_damp}). This advantage erodes as DV damping grows, leaving the CF-only $L=0$ limit most impervious.

	\subsection{Distinction from the \emph{bypass} protocol}
	\label{sec:comparison_bypass}

	The ``bypass'' protocol of Refs.~\cite{Park2022:Slowing, Hastrup2022:Universal} takes a different approach. It uses additional DV ancillas to transfer the quantum information to the DV system before the noise acts and to retrieve it (near-)deterministically afterwards. These bypass protocols, designed primarily for two- and four-component cat codes, are susceptible to DV damping noise and require additional gate resources and ancillas. Transferring CV quantum information into stationary DV systems, such as atoms in a cavity, also surrenders the practical advantages of traveling waves, including the ability to link distant quantum nodes.
	
\begin{figure}[htbp!]
		\centering
		\includegraphics[width=\columnwidth]{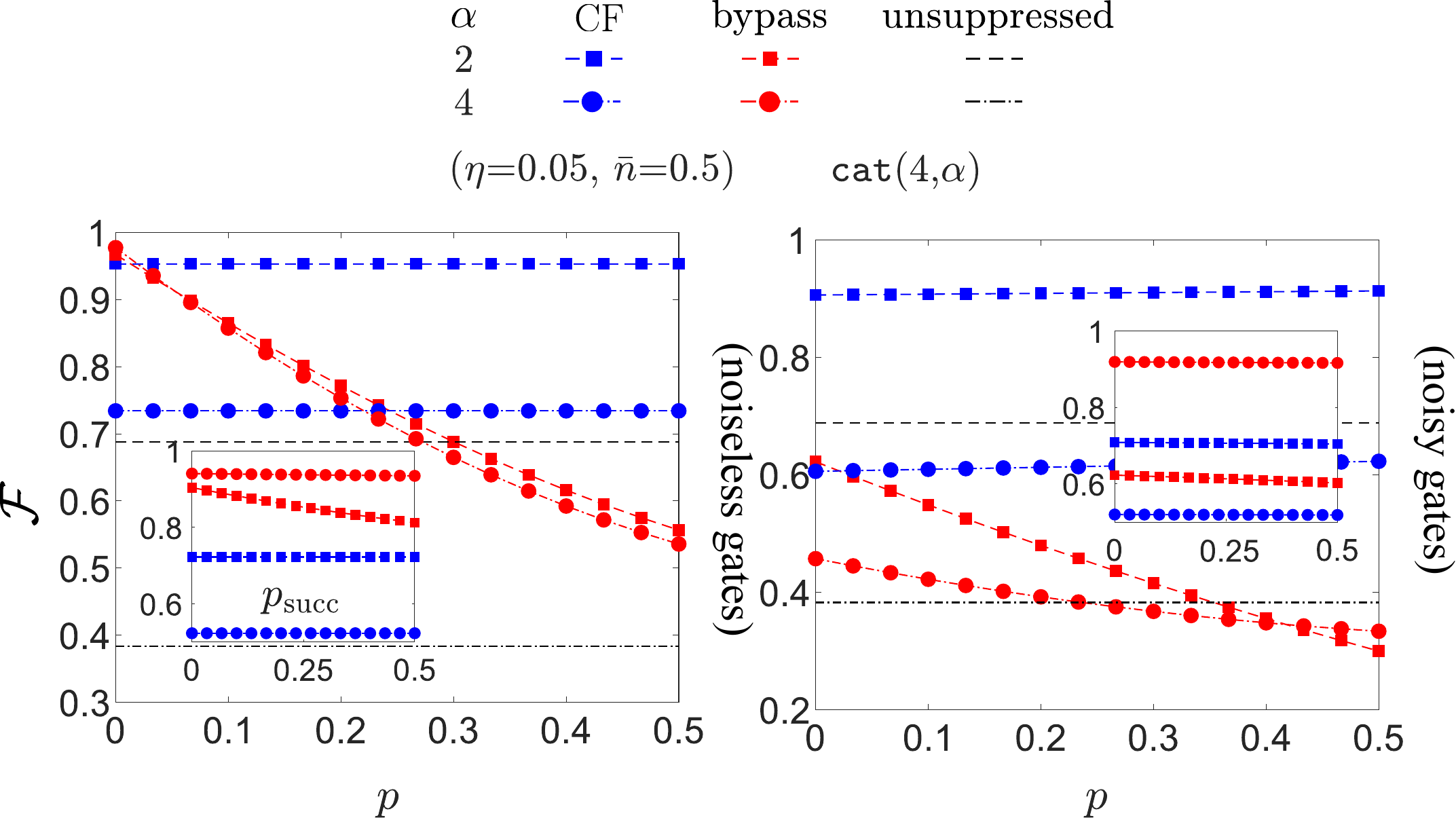}
		\caption{\label{fig:compare_bypass}
			Comparison between using conditional Fouriers and the conditional displacements prescribed by the ``bypass'' protocol for four-component cat codes, where the bypass protocol uses eight conditional displacements and two ancillas. The CF-only interferometer uses at most two conditional gates with a \mbox{single~ancilla}. The differences in the number of gates become important when the conditional gates are noisy. Here, all noisy entangling gates have an additional 1\% loss and composite DV damping rates. Insets show average success~probability.}
	\end{figure}

	Figures~\ref{fig:likevsopp} and~\ref{fig:PQPcondrot_damp} demonstrate that like-parity codes are impervious to DV damping noise because no state transfer from CV to DV occurs at any point, unlike in the bypass strategy. Even for opposite-parity codes, the CF-based approach with a single qubit ancilla is less susceptible to gate imperfections owing to its reduced gate count, as shown in Fig.~\ref{fig:compare_bypass}.
	
	\section{Qutrit-based noise-resilient suppression}
	\label{sec:qutrit}
	
	Extending beyond qubit ancillas, we demonstrate that a qumode subjected to thermal or Gaussian displacement noise can be protected using a qutrit ancilla and a pair of modified parity-detection primitives. This yields a truly hybrid CV-DV noise suppression scheme, remaining resilient even when the qutrit suffers from cascaded amplitude damping, $\ket{2} \to \ket{1} \to \ket{0}$, and phase damping of $\ketbra{1}{1}$.

	Qutrit ancillas are commonly used for noise-resilient measurements of photon number parity~\cite{Rosenblum2018:Fault-tolerant} and provide a good model for a noisy transmon. Parity detection $\pi$-rotates the qumode on the upper two qutrit eigenstates, with the central eigenstate flagging $\ket{2} \to \ket{1}$ damping.
	
	As in Sec.~\ref{sec:single_ancilla}, the suppressed CV thermal or Gaussian displacement noise Kraus operator pairs $A_l$, $B_k$, along with the qutrit-specific noise Kraus operator $K_j$ from Sec.~\ref{sec:noise_models}, are
	\begin{align}
		&\Us^\dagger K_j B_k A_l\Us =L_{l,k} \Us^\dagger(\AdA-l+k)K_j\Us(\AdA),
	\end{align}
	where the dependence of the suppression unitaries on the number operator $\AdA$ has been made explicit.
	The contribution from the DV ancilla is governed by $\Us^\dagger(\AdA-l+k)K_j\Us(\AdA)$, and is succinctly captured by the DV-only Choi--Jamio\l{}kowski operator using \mbox{the~isomorphism}
	\begin{equation}
		\VEC{\,\cdot\,}\equiv(\,\cdot\,_{\mathcal{H}_o}\otimes I_{\mathcal{H}_i})\ket{\Psi^{+}}\sqrt{d},
		\label{eq:CJ}
	\end{equation}
	where $\ket{\Psi^{+}}:=\sum_{j=1}^d\ket{j}\otimes\ket{j}/\sqrt{d}$ with an orthonormal set of states $\{\ket{j}\}$ is a maximally-entangled state between the $d$-dimensional Hilbert spaces $\mathcal{H}_o$ and $\mathcal{H}_i$. 
	Their matrix representations in the computational basis, referred to with the same notation henceforth, support a useful identity \mbox{$\VEC{ACB}=A\otimes B^\top\VEC{C}$}.

    \begin{figure}
		\centering
		\includegraphics[width=0.9\columnwidth]{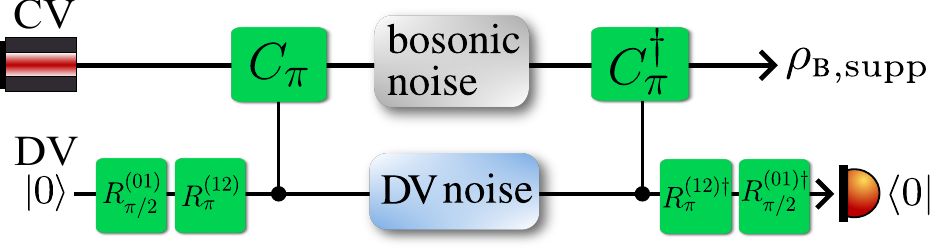}
		\caption{\label{fig:qutrit_protocol} Qutrit-based suppression using Pauli rotations embedded in the qutrit Hilbert space as well as a conditional-rotation of the qumode. The suppression unitary is a modification of the qutrit-based noise-resilient parity detector where the ancilla is first initiated in the $(\ket{0}+\ket{2})/\sqrt{2}$ state using the rotations $R_{\pi/2}^{(01)}$ and $R_{\pi}^{(12)}$, but the final $R^\dagger$s are removed. The complete setup pairs the modified qutrit-based parity detector with its adjoint to set up an interferometer as discussed earlier, leading to a truly hybrid, code-agnostic CV \emph{and} DV noise suppression.}
	\end{figure}
	\begin{figure}[h]
		\centering
		\includegraphics[width=\columnwidth]{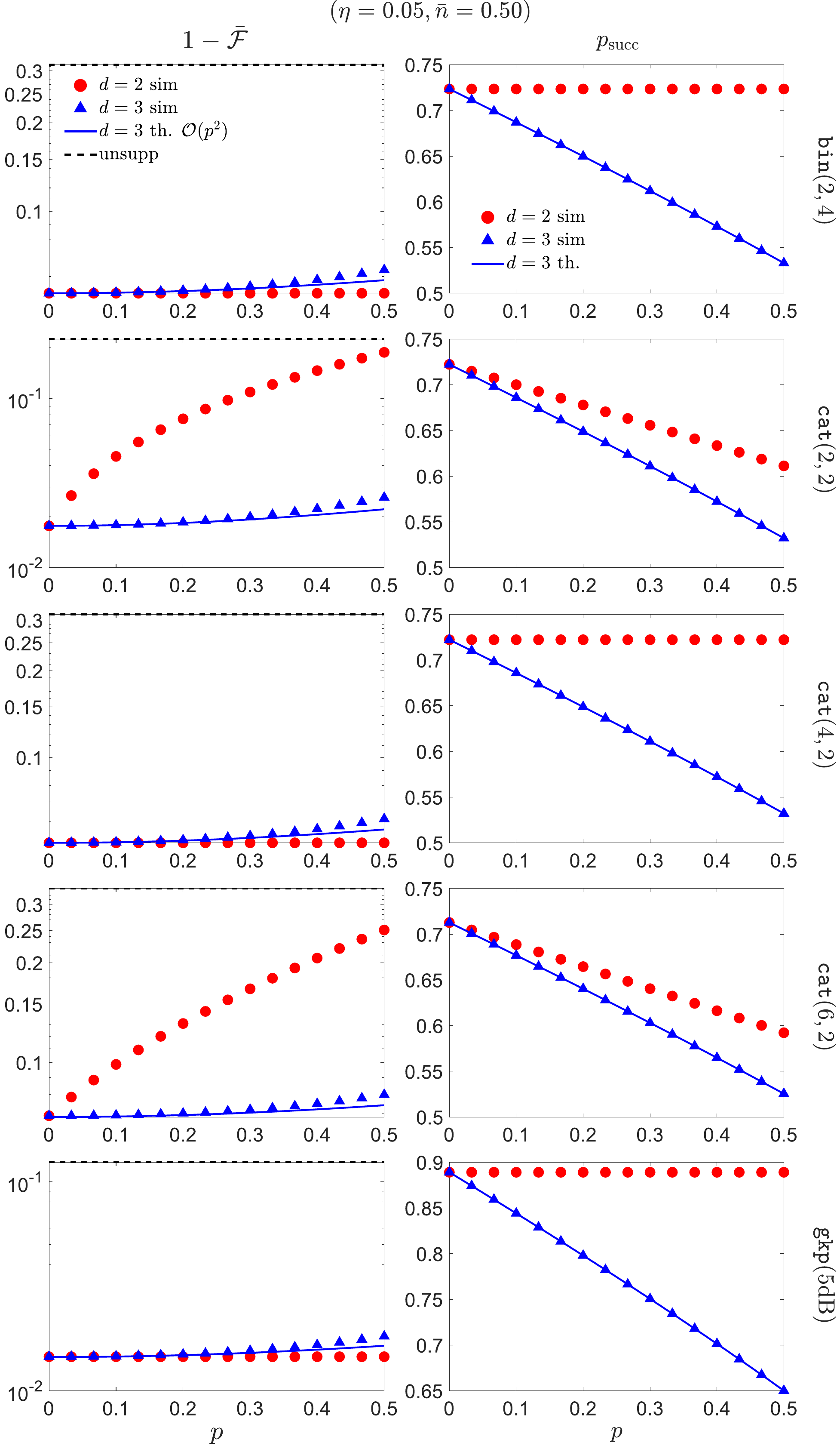}
		\caption{\label{fig:qutrit}
			Comparison between qubit-based ($d=2$) and qutrit-based ($d=3$) suppression performances for thermal noise with loss rate $\eta=0.05$ and a mean photon number $\bar{n}=0.5$, for various codes as a function of ancilla composite damping noise strength $p$. Both approaches suppress noise for all inputs, but a noise-resilient method with a qutrit is particularly beneficial for the opposite-parity codes.}
	\end{figure}
	
	Thus, the suppressed DV-Kraus operators are isomorphic to
	\begin{align}
		\Us^\dagger(\AdA-l+k)\otimes\Us^\top(\AdA)\VEC{K_j},
	\end{align}
	and they produce a Choi--Jamio\l{}kowski matrix 
	\begin{equation}
		J^{\textsc{q}}=\rho_{\textsc{B},n,m}\left(\Lambda_{n}J_0^{\textsc{q}}\Lambda_m^\dagger\right),
		\label{eq:partial-choi}
	\end{equation}
	for an arbitrary CV input defined by $\rhob := \sum_{n,m\geq 0} \ket{n}\rho_{\textsc{b},n,m}\bra{m}$
	where $\Lambda_n:=\Us^\dagger(n-l+k)\otimes\Us^\top(n)$ and \mbox{$J_0^{\textsc{q}}:=\sum_j\VEC{K_j}\!\BVEC{K_j}$} is the unsuppressed, DV-only, Choi--Jamio\l{}kowski matrix. The subscripts $k,l,m,n$ on $J^{\textsc{q}}$ have been omitted \mbox{for~convenience}. 
	
	This form immediately conveys how an arbitrary qumode input density matrix in the Fock basis is transformed by the interventions with the noisy DV ancilla.
	The complete CV-DV output state before any measurement becomes
	\begin{align}
		&\rho'_{\textsc{bq}}=\hspace{-1em} \sum_{k,l,m,n\geq 0}\hspace{-1em} 
		L_{l,k}\ket{n}
		\ptr{\mathcal{H}_i}{J^{\textsc{q}}(I_{\mathcal{H}_o}\otimes \rhoq^\top)}
		\bra{m}L_{l,k}^\dagger.
		\label{eq:choi-output}
	\end{align}
	The suppression unitary for this demonstration, depicted in Fig.~\ref{fig:qutrit_protocol}, is
	\begin{align}
		\Us(\AdA)=&\,\,C_{\pi}^{(12)}R_{\pi}^{(12)}R_{\pi/2}^{(01)}\notag\\
		\widehat{=}&\begin{pmatrix}
			1/\sqrt{2}  &   1/\sqrt{2}  & 0\\
			0           &   0           & \E{-\I \pi \AdA}\\
			\E{-\I \pi\AdA}/\sqrt{2}&-\E{-\I \pi \AdA}/\sqrt{2} & 0
		\end{pmatrix},
	\end{align}
	where $C_{\vartheta^{(jk)}}:=\E{-\I \vartheta\AdA\, \sigma_{0}^{(jk)}}$ is a conditional rotation of the CV on two qutrit eigenstates and similarly $R_{\vartheta}^{(jk)}:=\E{\I \frac{\vartheta}{2}\sigma_2^{(jk)}}$ is a Pauli rotation embedded in the qutrit's Hilbert space. The notation $\sigma_i^{(jk)}$ denotes Pauli-$i$ acting on the eigenstates $j$~and~$k$.
	
	For the qutrit ancilla initiated in $\rhoq=\ketbra{0}{0}$ and the composition of the noise channels given by Eqs.~\eqref{eq:qutrit-AD} and~\eqref{eq:qutrit-PD}, we observe immense simplification in Eq.~\eqref{eq:choi-output} whereby all the references to Fock states, $n$ and $m$, factor out of the DV contribution and 
	\begin{align}
		\ptr{\mathcal{H}_i}{J^{\textsc{q}}(I_{\mathcal{H}_o}\otimes \rhoq^\top)}
		=&\,\rho_{\textsc{b},n,m}D_{\textsc{q}},
		\label{eq:solved-choi}
	\end{align}
	where we defined
	\begin{align}
		D_{\textsc{q}}\,\,\widehat{=}\,\,\frac{1}{2}\begin{pmatrix}
			\substack{\displaystyle{(1-p/2)}\\\displaystyle{+(-1)^{l-k}\sqrt{1-p}}} & p/2 & 0\\
			p/2 & \substack{\displaystyle{(1-p/2)}\\\displaystyle{-(-1)^{l-k}\sqrt{1-p}}} & 0\\
			0 & 0 & p
		\end{pmatrix}.
	\end{align}
	Using Eq.~\eqref{eq:solved-choi} in Eq.~\eqref{eq:choi-output},
	\begin{align}
		\rho'_{\textsc{bq}}=\sum_{k,l\geq 0} 
		L_{l,k}\,\rhob\,L_{l,k}^\dagger\otimes
		D_{\textsc{q}},
		\label{eq:solved-output}
	\end{align}
	from which it is clear that the projection of the qutrit onto $\ketbra{2}{2}_{\textsc{q}}$ is a failure with no suppression,
	whereas the projections onto the state $\ketbra{j}{j}_{\textsc{q}}$ for $j\in\{0,1\}$ with $R_{l,k} := L_{l,k}\,\rhob\,L_{l,k}^\dagger$ lead to the \mbox{unnormalized~states}
	\begin{align}
		\tilde{\rho}_{\textsc{b},j}^\prime
		=&\,\sum_{k,l\geq 0}R_{l,k}\left[\frac{\left(1-p/2\right)}{2}+(-1)^{j+l-k}\frac{\sqrt{1-p}}{2}\right]\notag\\
		=&\,\frac{1-p/2+(-1)^j\sqrt{1-p}}{2}\sum_{\mathrm{even}(l-k)}\hspace{-1em}R_{l,k}\notag\\
		&+\frac{1-p/2+(-1)^{j+1}\sqrt{1-p}}{2}\sum_{\mathrm{odd}(l-k)}\hspace{-1em}R_{l,k}.
	\end{align} 
	Defining
\mbox{$\Delta:=\tr{\sum_{\mathrm{even}(l-k)}R_{l,k}-\sum_{\mathrm{odd}(l-k)}R_{l,k}}$}, 
		\begin{equation}
			\Sigma_0:=\hspace{-2em}\sum_{\substack{k,l\geq0\\(j+l+k)\bmod 2}}\hspace{-2em}R_{l,k}, \hspace{1em}\mathrm{and}\hspace{1em}  \Sigma_1:=\hspace{-2em}\sum_{\substack{k,l\geq0\\(1+j+l+k)\bmod 2}}\hspace{-2em}R_{l,k},
		\end{equation}
	the corresponding success probabilities read
	\begin{align}
		p_j=\frac{1-p/2}{2}+(-1)^j\frac{\sqrt{1-p}}{2}\Delta,
	\end{align}
	which for small $p$ furnish
	\begin{align}
		\tilde{\rho}_{\textsc{b},j}^\prime &\cong\left(1-\frac{p}{2}-\frac{p^2}{16}\right)\Sigma_0 +\frac{p^2}{16}\Sigma_1,\notag\\
		p_j^{-1} &\cong\frac{2}{1+(-1)^j\Delta}\notag\\
		&\quad+p\left(\frac{1}{1+(-1)^j\Delta}\right)+p^2\left(\frac{2+3(-1)^j\Delta}{4[1+(-1)^j\Delta]^2}\right),
	\end{align}
	which leads to the normalized states
	\begin{align}
		\rho'_{\textsc{b},j}\cong&\,\frac{2}{1+(-1)^j\Delta}\Bigg[\Sigma_0-\frac{p^2}{8[1+(-1)^j\Delta]}\notag\\
		&\,\times\Bigg(\frac{1-(-1)^j\Delta}{2}\Sigma_0-\frac{1+(-1)^j\Delta}{2}\Sigma_1\Bigg)\Bigg].
		\label{eq:final-state}
	\end{align}
	As the dominant term in the unnormalized state decays by a factor of $(1 - p/2)$, the terms $\propto p$ cancel perfectly upon normalization by the success probability, as seen in Eq.~\eqref{eq:final-state}. Moreover, for $j=0$, that is, by probabilistically projecting the ancilla back onto the state $\ketbra{0}{0}$, the dominant term $\Sigma_0$ will not contain the first-order terms of the CV noise as we have already seen for the qubit~case with Eq.~\eqref{eq:supp-state}. 
	
	Indeed, setting $p=0$ in Eq.~\eqref{eq:final-state} for $j=0$ exactly recovers the qubit CF gate result of Eq.~\eqref{eq:supp-state}, confirming that the qutrit protocol is a strict generalization of the single-qubit interferometer. Numerical investigation (not shown) also suggests that qutrit depolarizing noise can be suppressed effectively, owing to a smaller contribution from the \mbox{$\ket{0} \leftrightarrow \ket{2}$ transition}.
	
	\section{Conclusions and outlook}	
	\label{sec:conclusions}
	
	We analyzed a noise suppression protocol that targets photon-loss and thermal-noise corruption of single-mode bosonic systems or qumodes. It uses the hybrid entangling operations between the qumode and a single qubit ancilla, specifically the conditional Fourier gates sandwiching the noise, along with a final nondeterministic projection of the DV ancilla onto its original state. 
	
	Potential applications lie in quantum computing and communication, as the protocol addresses photon losses and thermal noise, which are the dominant noise sources for qumodes. Linear-optical mitigation methods using probabilistic error cancellation inevitably require measurements that reduce the bosonic-mode quantum state to classical numbers, and they also suffer from large sampling costs. Our noise suppression method retains the superposition state for further \emph{quantum} information processing, with a high success~rate.
	
	In the context of quantum computing with bosonic codes encoding qubit~information, we show that codes with orthogonal codewords of the like-parity type (whether odd or even) are completely impervious to additional ancilla damping noise, unlike those having opposite-parity codewords. Here, we presented a simple setup that uses at most two conditional Fourier gates. Similar behavior holds for $2^K$-fold rotation-symmetric bosonic codes under a multi-ancilla extension of the protocol. For such special codes, our protocol reduces to conventional error detection. A qutrit-based modification to the proposed method extends DV-damping-resilient CV noise suppression to codes that lack a photon-number parity syndrome, where conventional error correction does \mbox{not apply}.
	
	We also showed that by using multiple ancillas, one can convert the thermal noise channel into a form that is diagonal in the Fock basis. Combined with earlier linear-optical noise suppression protocols, our scheme can now convert thermal, random-displacement, and dephasing noise on a single bosonic mode into such a diagonal form. Going forward, three open problems are the noiseless amplification to counteract coherent damping, residual decoherence due to the mixture of Fock-diagonal noise, and state preparation and measurement (SPAM) imperfections of the DV ancilla outside the \mbox{suppression passage.}
	
	Further investigation of the error correction performances of concatenated CV-DV codes when supplemented with feasible noise suppression protocols, such as the one presented here, is the natural route to take. Bosonic codes that are easier to realize, but less capable in terms of error-correcting performance, may become competitive candidates after such noise-robust hybrid suppression~protocols. 
	This trade-off can be explored \mbox{in the future}.
	
	\section{Acknowledgments}
	The authors thank M.S. Kim, Y. Kim, K. Park, R. Filip, S. Bose and L. Jiang for insightful discussions. This work was supported by the Korean government [Ministry of Science and ICT (MSIT)], the NRF grants funded by the Korea government (MSIT) (No. RS-2023-00237959, No. RS-2024-00413957, No. RS-2024-00438415, No. RS-2025-02219034, No. RS-2025-25464492 and No. RS-2023-NR076733), the Institute of Information \& Communications Technology Planning \& Evaluation (IITP) grant funded by the Korea government (MSIT) (IITP-2025-RS-2020-II201606 and IITP-2025-RS-2024-00437191), and the Institute of Applied Physics at Seoul National University.
	HK is supported by the KIAS Individual Grant No. CG085302 at Korea Institute for Advanced Study. YST is supported by the faculty research fund of Sejong University in 2026.

	\newpage 
	\onecolumngrid
	\appendix
	
	{\allowdisplaybreaks
		\section{Success probability}
		\label{app:psucc_singlemodesupp}
		
		Here we analyze the success probability of single mode thermal or Gaussian random displacement noise suppression using multiple ancilla from the main text, given as	
		\begin{align}
			p_{\rm{succ}}=\mathop{\mathrm{tr}}\bigg\{\frac{\rhob}{G}x^{\AdA}\sum_{l,k\geq0}\frac{y^l}{l!}a^{\dagger l}\left[\frac{z^k}{k!}\,a^ka^{\dagger k}\delta_{(l-k)\bmod 2^K}\right]a^l 
			\bigg\},
			\label{eq:rho_output_unnormalized}
		\end{align}
		with definitions of $x$, $y$ and $z$ taken from the main text.
		
		The discrete Fourier transform of the Kronecker delta can be used to analytically pair up photon losses ($l$) and gains ($k$). For instance, with a single ancilla, we pair either the even or the odd $l$s and $k$s. 
		
		In general, the success probability is
		\begin{equation}
			p_{\rm{succ}}=\mathop{\mathrm{tr}}\bigg\{\frac{\rhob}{G} x^{\AdA}\sum_{l,k\geq0}\frac{y^l}{l!}
			a^{\dagger l}\bigg[\frac{z^k}{k!}\vdots(\AdA)^k\vdots \sum_{j=0}^{2^K-1}\frac{\E{\I\frac{2\pi}{2^K}(l-k) j}}{2^K}\bigg]a^l\bigg\}.
		\end{equation}
		We use $:\cdot:$ and $\vdots\cdot\vdots$ to denote normal and antinormal ordering of the bosonic ladder operators, respectively.
		By defining $\omega:=\E{\I {\pi}/{2^{K-1}}}$ to represent the relevant root of unity and collapsing the sum over $k$s into an exponential operator, it becomes
		\begin{equation}
			p_{\rm{succ}}=\frac{1-z}{2^K}\mathop{\mathrm{tr}}\bigg\{\rhob\,x^{\AdA}\sum_{j=0}^{2^K-1}\sum_{l\geq0}\frac{(y\omega^j)^l}{l!}
			a^{\dagger l} \vdots\E{z\omega^{- j}\AdA}\vdots\,a^l\bigg\}.
		\end{equation}
		The relations between operators under different orderings of the ladder operators (see Appendix~\ref{app:ordering-forms}) give
		\begin{align}
			p_{\rm{succ}}=\frac{1-z}{2^K}\mathop{\mathrm{tr}}\bigg\{\rhob\,x^{\AdA}\sum_{j=0}^{2^K-1}:
			\frac{\E{\big(y\omega^j+\frac{z\omega^{- j}}{1-z\omega^{-j}}\big)\AdA}}{1-z\omega^{-j}}:\bigg\}.
		\end{align}
		
		In case of a single ancilla, we have
		\begin{align}
			p_{\mathrm{succ}}=\frac{1-z}{2}\mathop{\mathrm{tr}}\bigg\{\rho\,x^{\AdA}\bigg[
			\frac{\left(1+y+\frac{z}{1-z}\right)^{\AdA}}{1-z}+\frac{\bigg(1-y-\frac{z}{1+z}\bigg)^{\AdA}}{1+z}\bigg]\bigg\}=\frac{1}{2}+\frac{1}{2(2G-1)}\mathop{\mathrm{tr}}\bigg\{\rho
			\bigg(\frac{1-2\mu G}{2G-1}\bigg)^{\AdA}\bigg\}.
			\label{eq:psucc_1dv}
		\end{align}
		whereas in the asymptotic limit ($K\rightarrow\infty$)
		\begin{align}
			p_{\mathrm{succ}}&=(1-z)\mathop{\mathrm{tr}}\bigg\{\rho\,x^{\AdA}\sum_{l\geq0}\frac{y^l}{l!}a^{\dagger l}\left[\sum_{k\geq 0}\frac{z^k}{k!}\,a^ka^{\dagger k}\delta_{(l-k)}\right]a^l\bigg\} \notag\\
			&=(1-z)\mathop{\mathrm{tr}}\bigg\{\rho\,x^{\AdA}\sum_{l\geq0}\frac{(yz)^l}{(l!)^2}a^{\dagger l}a^l\,a^{\dagger l}a^l\bigg\}\notag\\
			&=(1-z)\mathop{\mathrm{tr}}\bigg\{\rho\,x^{\AdA}\sum_{n,l\geq0}\frac{(yz)^l}{(l!)^2}\left(\frac{n!}{(n-l)!}\right)^2\ketbra{n}{n}\bigg\}\notag\\
			&=(1-z)\mathop{\mathrm{tr}}\bigg\{\rho\,x^{\AdA}\sum_{n,l\geq0}(yz)^l\,{{n}\choose{l}}^2\ketbra{n}{n}\bigg\}\notag\\
			&=(1-z)\mathop{\mathrm{tr}}\bigg\{\rho\,x^{\AdA}\sum_{n\geq0}(1-yz)^n\,P_{n}\left(\frac{1+yz}{1-yz}\right) \ketbra{n}{n}\bigg\}\notag\\
			&=(1-z)\mathop{\mathrm{tr}}\bigg\{\rho\,[x(1-yz)]^{\AdA}\,P_{\AdA}\left(\frac{1+yz}{1-yz}\right)\bigg\}\notag\\
			&=\frac{1}{G}\mathop{\mathrm{tr}}\bigg\{\rho \bigg(\frac{1}{G}-\mu\bigg)^{\AdA}P_{\AdA}\bigg(\frac{1-\mu(2-G)}{1-\mu G}\bigg)\bigg\},
		\end{align}
		where $P_n(\cdot)$ are the Legendre polynomials.
		
		\section{Linear order suppression of physical noise in fidelity}
		\label{app:linear_order_suppression}
		
		This appendix continues the details and calculations outlined in the main text, showing linear order suppression. 
		For a noiseless DV ancilla, we obtain 
		$\tilde{\rho}'=\sum_{j,k}P_{j,k}z^j\mu^k$
		such that
		\begin{align}
			&P_{0,0}:=\rho, P_{1,0}:=-(\rho+\{T_1,\rho\}), P_{0,1}:=-\{T_1,\rho\}\notag\\
			&P_{1,1}:=[\rho,T_1]+\{T_{2,1},\rho\}-2T_1\rho\,T_1+\AdA\rho\,\AdA\notag\\
			&P_{2,0}:=[\rho,T_1]+\{T_{2,2},\rho\}-T_1\rho\,T_1+\frac{1}{2}a^{\dagger\,2}\rho\,a^2\notag\\
			&P_{0,2}:=\{\rho,T_{2,2}\}-T_1\rho\,T_1+\frac{1}{2}a^2\rho\,a^{\dagger\,2}.
		\end{align}

		The normalization of this state leads to the success probability $p_{\mathrm{succ}}\cong\sum_{j,k}C_{j,k}z^j\mu^k$ such that
		\begin{align}
			&C_{0,0}:=1, C_{1,0}=-(1+2\MEAN{T_1}), C_{0,1}:=-2\MEAN{T_1}\notag\\
			&C_{1,1}:=2\MEAN{T_{2,1}}-2\MEAN{T_1^2}+\MEAN{(\AdA)^2}\notag\\
			&C_{2,0}:=2\MEAN{T_{2,2}}-\MEAN{T_{2,1}^2}+\frac{1}{2}\big\langle{\vdots(\AdA)^2\vdots}\big\rangle\notag\\
			&C_{0,2}:=2\MEAN{T_{2,2}}-\MEAN{T_{2,1}^2}+\frac{1}{2}\MEAN{:(\AdA)^2:},
			\label{eq:psucc_coef}
		\end{align}
		where we employ $\MEAN{\cdot}$ for the expectation values in the state $\rho$
		\begin{align}
			a^{\dagger 2}a^2=&\AdA (\AdA-1),\notag\\
			a^{2}a^{\dagger 2}=&(\AdA+1)(\AdA+2), 
		\end{align}
		using the commutators of the bosonic ladder operators.
		
		Now, we expand the inverse of the success probability to the second order in $\mu$ and $z$ giving $p^{-1}_{\mathrm{succ}}\cong\sum_{j,k}Q_{j,k}z^j\mu^k$ such that
		\begin{align}
			&Q_{0,0}:=\frac{1}{C_{0,0}}, Q_{1,0}:=-\frac{C_{1,0}}{C_{0,0}^2}, Q_{0,1}:=-\frac{C_{0,1}}{C_{0,0}^2},\notag\\
			&Q_{1,1}:=\frac{2C_{0,1}C_{1,0}-C_{0,0}C_{1,1}}{C_{0,0}^3},\notag\\
			&Q_{2,0}:=\frac{C_{1,0}^2-C_{0,0}C_{2,0}}{C_{0,0}^3},\notag\\
			&Q_{0,2}:=\frac{C_{0,1}^2-C_{0,0}C_{0,2}}{C_{0,0}^3}.
			\label{eq:psucc_inv_coef}
		\end{align}
		
		The normalized output state is therefore $\rho'=\sum_{j,k}P_{j,k}Q_{l,m}z^{j+l}\mu^{k+m}$ such that
		\begin{align}
			&R_{0,0}:=\rho,\notag\\
			&R_{1,0}=R_{0,1}:=2\MEAN{T_1}\rho-\{T_1,\rho\},\notag\\
			&R_{1,1}:=[\rho,T_1]+\{\rho,T_{2,1}\}-2\,T_1\rho\,T_1+\AdA\rho\,\AdA\notag\\
			&+\rho\left[4\MEAN{T_1}(2\MEAN{T_1}+1)+2\MEAN{T_1^2}-2\MEAN{T_{2,1}}-\MEAN{(\AdA)^2}\right]\notag\\
			&-2\MEAN{T_1}\rho-4\MEAN{T_1}\{T_1,\rho\}\notag\\
			&R_{2,0}:=[\rho,T_{1}]+\{\rho,T_{2,2}\}-T_1\rho\,T_1+\frac{1}{2}a^{\dagger\,2}\rho\,a^2\notag\\
			&+\rho\bigg[(2\MEAN{T_1}+1)^2+\MEAN{T_1^2}-2\MEAN{T_{2,2}}-\frac{1}{2}\MEAN{\vdots(\AdA)^2\vdots}\bigg]\notag\\
			&R_{0,2}:=\{\rho,T_{2,2}\}-T_1\rho\,T_1+\frac{1}{2}a^{2}\rho\,a^{\dagger\,2}\notag\\
			&+\rho\bigg[4\MEAN{T_1}^2+\MEAN{T_1^2}-2\MEAN{T_{2,2}}-\frac{1}{2}\MEAN{:(\AdA)^2:}\bigg]
		\end{align}
		The normalized state is therefore 
		\begin{align}
			\rho'&\cong\rho+z\big(2\MEAN{T_1}\rho-\{T_1,\rho\}\big)+\mu\big(2\MEAN{T_1}\rho-\{T_1,\rho\}\big)+\mu z R_{1,1}+z^2\,R_{2,0}+\mu^2\,R_{0,2}.
		\end{align}
		
		\section{Average fidelity under thermal noise without suppression}
		\label{app:Funsupp}
		
		Without any noise suppression, Eq.~\eqref{eq:supp-state} doesn't possess the final DV term and therefore, the pairings $(k,l)=(1,0)$ and $(0,1)$ do not vanish.
		
		Consequently, the (normalized) state is
		\begin{equation}
			\rho'\cong\rho-\eta\args{\bar{n}\rho+\argc{\argp{\bar{n}+\frac{1}{2}}\AdA,\rho} -\bar{n}a^\dagger\,\rho\,a -(1+\bar{n})a\,\rho\,a^\dagger},
		\end{equation}
		giving the fidelity,
		\begin{align}
			\tr{\rho \rho'}\cong&1-\eta\args{\bar{n}+(1+2\bar{n})\MEAN{\AdA}-\bar{n}\tr{\rho\,a^\dagger\,\rho\,a}-(1+\bar{n})\tr{\rho\,a\,\rho\,a^\dagger}}.
		\end{align}
		
		The above equation is averaged to
		\begin{equation}
			\overline{\mathcal{F}}_{\mathrm{unsupp}}\cong1-\eta\args{\bar{n}+(1+2\bar{n})\tr{\SL\,\AdA} -\frac{2\bar{n}}{3}\argp{\tr{\SL\,a^\dagger\,\SL\,a}+|\tr{\SL\,a}|^2}-\frac{2(1+\bar{n})}{3}\argp{\tr{\SL\,a\SL\,a^\dagger}+|\tr{\SL\,a}|^2}},
		\end{equation}
		where we now see the undesirable linear scaling with $\eta$. Moreover, note that the final two terms in the parentheses vanish for the like parity codes, suggesting they are affected more severely than their opposite parity counterparts with the identical mean photon number ($\tr{\SL\,\AdA}$).
		
		\section{Application: Noise suppression for quantum communication}
		\label{sec:comm}
		
		While the main text focuses on localized suppression, controlled rotations on traveling waves are exceptionally suited for non-local quantum communication~\cite{Munro2012:Quantum, Li2023:Memoryless, Li2024:Performance}. Utilizing preshared DV entanglement between remote nodes alongside classical communication~\cite{Ritter2012:Elementary} (Fig.~\ref{fig:comm}~(a),(b)), the protocol suppresses transmission noise entirely on the fly. For perfect ancillas and noiseless CF gates, $\overline{\mathcal{F}}_\mathrm{supp}$ and $p_\mathrm{succ}$ exactly match the localized protocol (Eqs.~\eqref{eq:Fsupp} and~\eqref{eq:psucc}).
		
		This architecture utilizes stationary DV systems as resource memories, while traveling qumodes facilitate the actual communication. Furthermore, unlike standard DV teleportation, all measurements are fixed projections requiring no active feedforward. 	
		
		\begin{figure}[h]
			\centering
			\includegraphics[width=0.5\columnwidth]{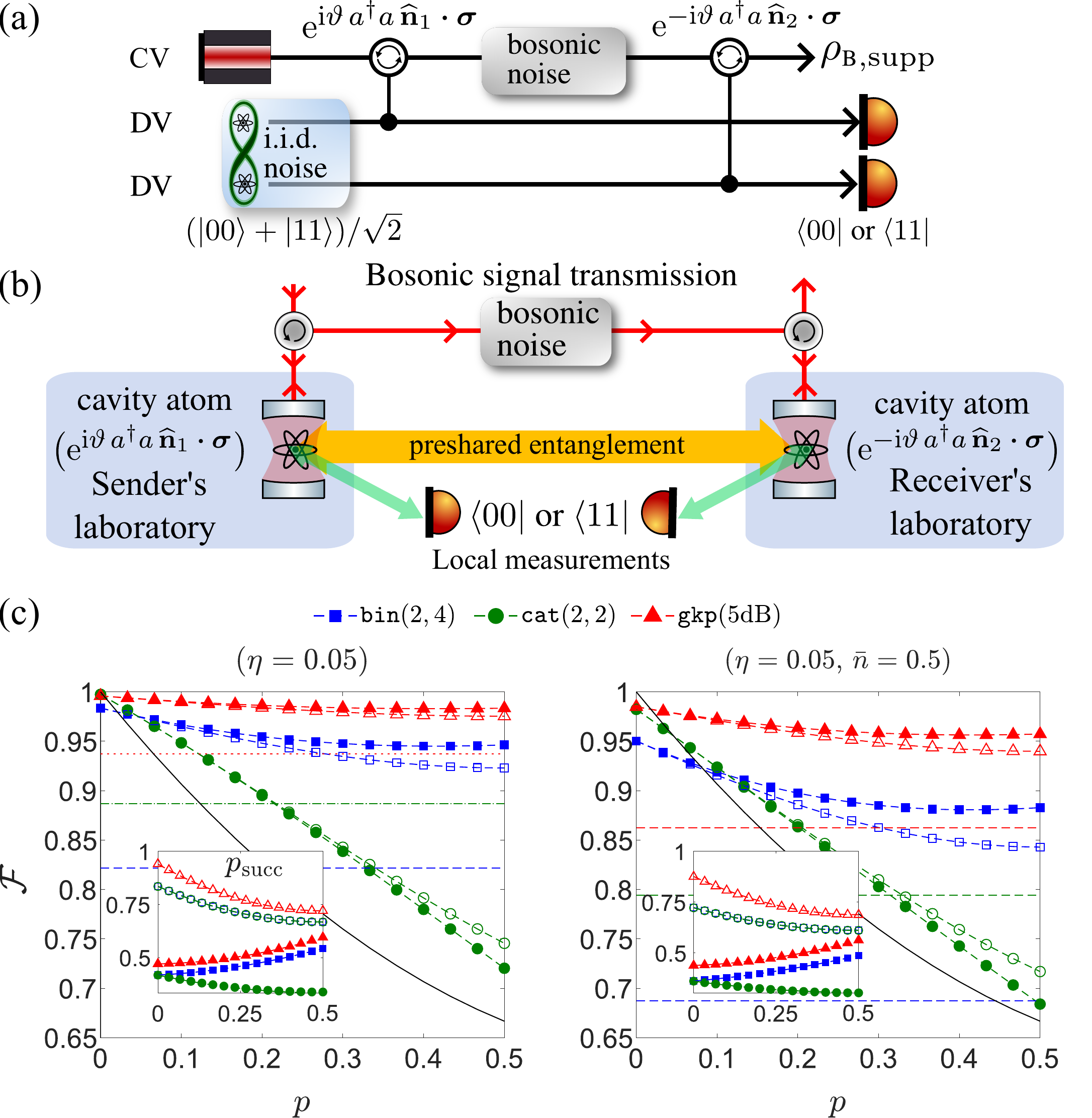}
			\caption{\label{fig:comm} 
				The like parity codes are also resilient to the composite amplitude and phase damping ancilla noise in quantum communication.
				(a) Circuit for bosonic signal transmission with preshared DV entanglement and classical communications as a resource. (b) An optical circuit to implement the protocol for quantum signal transmission using two remote cavity atom systems. (c) The average performance of various encodings under the noise suppression protocol for both photon loss of rate $\eta=0.05$ and thermal noise of the same rate and $\bar{n}=0.5$. The unmarked dashed, dot-dashed, and dotted lines refer to the average unsuppressed fidelities for the binomial, cat, and GKP codes, while the solid line is the average DV teleportation fidelity. Insets show average success~probability. The filled and unfilled markers refer to heralding the bosonic output on the ancillary measurement outcome~$00$, and both outcomes $00$ and $11$,~respectively.}
		\end{figure}
		
		Results for various bosonic codes in Fig.~\ref{fig:comm} show the advantage of our protocol in the presence of composite DV damping noise on the resource \mbox{Bell~state}.
		The average DV-only teleportation fidelity assuming that only composite amplitude and phase damping noise acts on the shared Bell state is given by $\overline{\mathcal{F}}_\mathrm{tele}=1-p+2p^2/3$ with damping parameter $0 \leq p\leq 1$
		(see Appendix~\ref{app:DV_tele_fid} or \cite{Horodecki1996:Teleportation}) and the figure reveals that teleportation requires high-quality DV Bell states, whereas our scheme is resilient to damping~noise. Moreover, the DV-only protocol does not support further error correction.
		
		In the presence of composite damping noise, the performance in terms of the average success probability of this protocol is given by (see Appendix~\ref{app:psucc_CVcomm}),
		\begin{equation}
			\overline{p}_\mathrm{succ}=\frac{1}{2}+\frac{1-2p(1-p)}{2(2G-1)}\tr{\SL
				\left(\frac{1-2\mu G}{2G-1}\right)^{\AdA}},
			\label{eq:psucc_00_or_11}
		\end{equation}
		when the output is heralded on \emph{both} $00$ and $11$~outcomes. With damping noise, the average fidelity of the even-parity codes may be slightly enhanced by only heralding on~$00$ as shown by the filled markers in Fig.~\ref{fig:comm}, at a reduced success~probability.			
		
		\section{Success probability of quantum communication setup under thermal noise and ancilla damping noise}
		\label{app:psucc_CVcomm}
		
		The setup for bosonic signal transmission requires a preshared entanglement stored in the remote atoms in the form of a Bell state $\ket{\Phi^+}=\big(\ket{00}+\ket{11}\big)/\sqrt{2}$. Here, we consider the effect of the composite DV damping on the Bell state ancilla. The independent and identically distributed (i.i.d.) damping noise channel over two qubits can be constructed using  Eq.~\eqref{eq:damp} on the various terms of the density operator $\rho_{\Phi^+}=\ketbra{\Phi^+}{\Phi^+}$ as
		\begin{align}
			\mathcal{N}_{\text{damp}}^{\otimes 2}[\ketbra{00}{00}]=&\frac{(c_++c_-+2p)^2}{16}\ketbra{00}{00}=\ketbra{00}{00},\notag\\
			\mathcal{N}_{\text{damp}}^{\otimes 2}[\ketbra{11}{11}]=&\frac{(c_++c_--2p)^2}{16}\ketbra{11}{11}+p^2\ketbra{00}{00}+\frac{p(c_++c_--2p)}{4}\big(\ketbra{01}{01}+\ketbra{10}{10}\big),\notag\\
			=&(1-p)^2\ketbra{11}{11}+p^2\ketbra{00}{00}+p(1-p)\big(\ketbra{01}{01}+\ketbra{10}{10}\big),\notag\\
			\mathcal{N}_{\text{damp}}^{\otimes 2}[\ket{00}\!\!\bra{11}]=&\frac{(c_+-c_-)^2}{16}\ket{00}\!\!\bra{11}=(1-p)^2\ket{00}\!\!\bra{11},\notag\\
			\mathcal{N}_{\text{damp}}^{\otimes 2}[\ket{11}\!\!\bra{00}]=&\frac{(c_+-c_-)^2}{16}\ket{11}\!\!\bra{00}=(1-p)^2\ket{11}\!\!\bra{00},
		\end{align}
		which leads to
		\begin{align}
			\mathcal{N}_{\text{damp}}^{\otimes 2}[\rho_{\Phi^+}]=&\frac{1}{2}\bigg\{
			(1+p^2)\ketbra{00}{00}+(1-p)^2\ketbra{11}{11}+p(1-p)\big(\ketbra{01}{01}+\ketbra{10}{10}\big)
			+(1-p)^2\,\big(\ket{00}\!\!\bra{11}+\ket{11}\!\!\bra{00}\big)
			\bigg\}.
			\label{eq:noisy-Bell}
		\end{align}
		
		Due to the linearity of transformations, we find the relevant inner products of the form $K_{s}(\mu,\nu,l,k):=\opinner{}{\Us^{(2)} B_kA_l \Us^{(1)}}{\mu\nu}$, where $\bra{}$ denotes the conditioned outcomes (either $\bra{00}$ or $\bra{11}$) and $\ket{\mu\nu}$ is ket representing the state of two DV ancilla with $\mu,\nu=$ 0 or 1, $A_l$ and $B_k$ are the $l$th and the $k$th Kraus operators of loss and quantum-limited amplification channels respectively, and $\Us^{(1)}:=\exp\big(\I \frac{\pi}{2}\AdA\, \sigma_1^{(1)}\big)$ and $\Us^{(2)}:=\exp\big(-\I \frac{\pi}{2}\AdA\, \sigma_1^{(2)}\big)$ are the unitaries implemented in the sender's and receiver's laboratory respectively. These inner products can be viewed as suppressed Kraus operators, which may not be complete and trace-preserving.
		
		Now, we follow the procedure similar to obtaining Eq.~(\ref{eq:psucc_1dv}). First, by dropping the parameters in the notation for simplicity, we have the suppressed Kraus operators
		\begin{align}
			&K_s=c_{l,k}\sqrt{x}^{\AdA}\,\opinner{}{\E{-\I \frac{\pi}{2}\AdA \sigma_1^{(2)}} a^{\dagger k}a^l\E{\I \frac{\pi}{2}\AdA \sigma_1^{(1)}}}{\mu\nu}\notag\\
			&=c_{l,k}\sqrt{x}^{\AdA}a^{\dagger k} a^l\,\big[\cos\argp{\frac{\pi}{2}(\AdA+k-l)}\cos\left(\frac{\pi}{2}\AdA\right)\inner{}{\mu\nu}+\I\cos\left(\frac{\pi}{2}(\AdA+k-l)\right)\sin\left(\frac{\pi}{2}\AdA\right)\opinner{}{\sigma_1^{(1)}}{\mu\nu}\notag\\
			&{\hskip8em\relax}-\I\sin\argp{\frac{\pi}{2}(\AdA+k-l)}\cos\left(\frac{\pi}{2}\AdA\right)\opinner{}{\sigma_1^{(2)}}{\mu\nu}+\sin\argp{\frac{\pi}{2}(\AdA+k-l)}\sin\left(\frac{\pi}{2}\AdA\right)\opinner{}{\sigma_1^{(2)}\sigma_1^{(1)}}{\mu\nu}\big].
			\label{eq:suppK}
		\end{align}
		
		Using these, we get the unnormalized density operators as
		\begin{align}
			\tilde{\rho}^\prime(00)=&\frac{1}{2}\sum_{l,k\geq0}c_{l,k}^2\sqrt{x}^{\AdA}a^{\dagger k}a^l\bigg\{(1+p^2)\,\cos\argp{\frac{\pi}{2}(\AdA+k-l)}\cos\left(\frac{\pi}{2}\AdA\right)\rho\cos\left(\frac{\pi}{2}\AdA\right)\cos\argp{\frac{\pi}{2}(\AdA+k-l)} \notag\\
			&+(1-p)^2\, \sin\argp{\frac{\pi}{2}(\AdA+k-l)}\sin\left(\frac{\pi}{2}\AdA\right)\rho\sin\left(\frac{\pi}{2}\AdA\right)\sin\argp{\frac{\pi}{2}(\AdA+k-l)} \notag\\
			&+p(1-p)\,\bigg[ \cos\argp{\frac{\pi}{2}(\AdA+k-l)}\sin\left(\frac{\pi}{2}\AdA\right)\rho\sin\left(\frac{\pi}{2}\AdA\right)\cos\argp{\frac{\pi}{2}(\AdA+k-l)}\notag\\
			&{\hskip6em\relax}+\sin\argp{\frac{\pi}{2}(\AdA+k-l)}\cos\left(\frac{\pi}{2}\AdA\right)\rho\cos\left(\frac{\pi}{2}\AdA\right)\sin\argp{\frac{\pi}{2}(\AdA+k-l)}
			\bigg]\notag\\
			&+(1-p)^2\,\bigg[ \cos\argp{\frac{\pi}{2}(\AdA+k-l)}\cos\left(\frac{\pi}{2}\AdA\right)\rho\sin\left(\frac{\pi}{2}\AdA\right)\sin\argp{\frac{\pi}{2}(\AdA+k-l)}\notag\\
			&{\hskip6em\relax}+\sin\argp{\frac{\pi}{2}(\AdA+k-l)}\sin\left(\frac{\pi}{2}\AdA\right)\rho\cos\left(\frac{\pi}{2}\AdA\right)\cos\argp{\frac{\pi}{2}(\AdA+k-l)}
			\bigg]
			\bigg\}a^{\dagger l}a^k\sqrt{x}^{\AdA}.
		\end{align}
		and similarly,
		\begin{align}
			\tilde{\rho}^\prime(11)=&\frac{1}{2}\sum_{l,k\geq0}c_{l,k}^2\sqrt{x}^{\AdA}a^{\dagger k}a^l\bigg\{(1-p)^2\,\cos\argp{\frac{\pi}{2}(\AdA+k-l)}\cos\left(\frac{\pi}{2}\AdA\right)\rho\cos\left(\frac{\pi}{2}\AdA\right)\cos\argp{\frac{\pi}{2}(\AdA+k-l)} \notag\\
			&+(1+p^2)\, \sin\argp{\frac{\pi}{2}(\AdA+k-l)}\sin\left(\frac{\pi}{2}\AdA\right)\rho\sin\left(\frac{\pi}{2}\AdA\right)\sin\argp{\frac{\pi}{2}(\AdA+k-l)} \notag\\
			&+p(1-p)\,\bigg[ \cos\argp{\frac{\pi}{2}(\AdA+k-l)}\sin\left(\frac{\pi}{2}\AdA\right)\rho\sin\left(\frac{\pi}{2}\AdA\right)\cos\argp{\frac{\pi}{2}(\AdA+k-l)}\notag\\
			&{\hskip6em\relax}+\sin\argp{\frac{\pi}{2}(\AdA+k-l)}\cos\left(\frac{\pi}{2}\AdA\right)\rho\cos\left(\frac{\pi}{2}\AdA\right)\sin\argp{\frac{\pi}{2}(\AdA+k-l)}
			\bigg]\notag\\
			&+(1-p)^2\,\bigg[ \cos\argp{\frac{\pi}{2}(\AdA+k-l)}\cos\left(\frac{\pi}{2}\AdA\right)\rho\sin\left(\frac{\pi}{2}\AdA\right)\sin\argp{\frac{\pi}{2}(\AdA+k-l)}\notag\\
			&{\hskip6em\relax}+\sin\argp{\frac{\pi}{2}(\AdA+k-l)}\sin\left(\frac{\pi}{2}\AdA\right)\rho\cos\left(\frac{\pi}{2}\AdA\right)\cos\argp{\frac{\pi}{2}(\AdA+k-l)}
			\bigg]
			\bigg\}a^{\dagger l}a^k\sqrt{x}^{\AdA},
		\end{align}
		for the two outcomes of interest, $00$ and $11$ respectively.
		
		The probabilities of obtaining the state above are their traces. Therefore, for the outcome $00$ we have
		\begin{align}
			&\,\tr{\tilde{\rho}^\prime(00)}\notag\\
			&=\frac{(1+p^2)}{2G}\tr{\cos\left(\frac{\pi}{2}\AdA\right)\rho\cos\left(\frac{\pi}{2}\AdA\right)x^{\AdA}\sum_{l\geq0}\frac{y^l}{l!}\sum_{k\geq0}\frac{z^k}{k!}\cos\argp{\frac{\pi}{2}(\AdA+k-l)}a^{\dagger l}\vdots (\AdA)^k\vdots\,a^l\cos\argp{\frac{\pi}{2}(\AdA+k-l)}}\notag\\
			&+\frac{(1-p)^2}{2G}\tr{\sin\left(\frac{\pi}{2}\AdA\right)\rho\sin\left(\frac{\pi}{2}\AdA\right)x^{\AdA}\sum_{l\geq0}\frac{y^l}{l!}\sum_{k\geq0}\frac{z^k}{k!}\sin\argp{\frac{\pi}{2}(\AdA+k-l)}a^{\dagger l}\vdots (\AdA)^k\vdots\,a^l\sin\argp{\frac{\pi}{2}(\AdA+k-l)}}\notag\\
			&+\frac{p(1-p)}{2G}\bigg[\tr{\sin\left(\frac{\pi}{2}\AdA\right)\rho\sin\left(\frac{\pi}{2}\AdA\right)x^{\AdA}\sum_{l\geq0}\frac{y^l}{l!}\sum_{k\geq0}\frac{z^k}{k!}\cos\argp{\frac{\pi}{2}(\AdA+k-l)}a^{\dagger l}\vdots (\AdA)^k\vdots\,a^l\cos\argp{\frac{\pi}{2}(\AdA+k-l)}}\notag\\
			&{\hskip4em\relax}+\tr{\cos\left(\frac{\pi}{2}\AdA\right)\rho\cos\left(\frac{\pi}{2}\AdA\right)x^{\AdA}\sum_{l\geq0}\frac{y^l}{l!}\sum_{k\geq0}\frac{z^k}{k!}\sin\argp{\frac{\pi}{2}(\AdA+k-l)}a^{\dagger l}\vdots (\AdA)^k\vdots\,a^l\sin\argp{\frac{\pi}{2}(\AdA+k-l)}}\bigg]\notag\\
			&+\frac{(1-p)^2}{2G}\bigg[\tr{\sin\left(\frac{\pi}{2}\AdA\right)\rho\cos\left(\frac{\pi}{2}\AdA\right)x^{\AdA}\sum_{l\geq0}\frac{y^l}{l!}\sum_{k\geq0}\frac{z^k}{k!}\cos\argp{\frac{\pi}{2}(\AdA+k-l)}a^{\dagger l}\vdots (\AdA)^k\vdots\,a^l\sin\argp{\frac{\pi}{2}(\AdA+k-l)}}\notag\\
			&{\hskip4em\relax}+\tr{\cos\left(\frac{\pi}{2}\AdA\right)\rho\sin\left(\frac{\pi}{2}\AdA\right)x^{\AdA}\sum_{l\geq0}\frac{y^l}{l!}\sum_{k\geq0}\frac{z^k}{k!}\sin\argp{\frac{\pi}{2}(\AdA+k-l)}a^{\dagger l}\vdots (\AdA)^k\vdots\,a^l\cos\argp{\frac{\pi}{2}(\AdA+k-l)}}\bigg]
		\end{align}
		with $x, y, z$ as defined in Sec.~\ref{sec:single_ancilla}.
		
		The operators are separated from the constant factors using simple trigonometric identities, and $[F(\AdA),a^{\dagger \,l}\vdots(\AdA)^{k}\vdots\,a^l]=0$ for any function of $F(\AdA)$ is used to give
		\begin{align}
			&\,\tr{\tilde{\rho}^\prime(00)}\notag\\
			&=\frac{(1+p^2)}{2G}\mathop{\mathrm{tr}}\Bigg\{\cos\argp{\frac{\pi}{2}\AdA}\rho\cos\argp{\frac{\pi}{2}\AdA}x^{\AdA}\notag\\
			&{\hskip4em\relax}\sum_{l\geq0}\frac{y^l}{l!}\sum_{k\geq0}\frac{z^k}{k!}\args{\cos^2\argp{\frac{\pi}{2}\AdA}\delta_{\mathrm{even}\,(k-l)}+\sin^2\argp{\frac{\pi}{2}\AdA}\delta_{\mathrm{odd}\,(k-l)}-\frac{\cancelto{0}{\sin(\pi\AdA)}\cancelto{0}{\sin(\pi(k-l))}}{4}\quad}a^{\dagger \,l}\vdots(\AdA)^{k}\vdots\,a^l\Bigg\}\notag\\
			&+\frac{(1-p)^2}{2G}\mathop{\mathrm{tr}}\argc{\sin\argp{\frac{\pi}{2}\AdA}\rho\sin\argp{\frac{\pi}{2}\AdA}x^{\AdA}\sum_{l\geq0}\frac{y^l}{l!}\sum_{k\geq0}\frac{z^k}{k!}\args{\cos^2\argp{\frac{\pi}{2}\AdA}\delta_{\mathrm{odd}\,(k-l)}+\sin^2\argp{\frac{\pi}{2}\AdA}\delta_{\mathrm{even}\,(k-l)}}a^{\dagger \,l}\vdots(\AdA)^{k}\vdots\,a^l}\notag\\
			&+\frac{p(1-p)}{2G}\Bigg[\tr{\sin\argp{\frac{\pi}{2}\AdA}\rho\sin\argp{\frac{\pi}{2}\AdA}x^{\AdA}\sum_{l\geq0}\frac{y^l}{l!}\sum_{k\geq0}\frac{z^k}{k!}\args{\cos^2\argp{\frac{\pi}{2}\AdA}\delta_{\mathrm{even}\,(k-l)}+\sin^2\argp{\frac{\pi}{2}\AdA}\delta_{\mathrm{odd}\,(k-l)}}a^{\dagger \,l}\vdots(\AdA)^{k}\vdots\,a^l}\notag\\
			&{\hskip4em\relax}+\tr{\cos\left(\frac{\pi}{2}\AdA\right)\rho\cos\argp{\frac{\pi}{2}\AdA}x^{\AdA}\sum_{l\geq0}\frac{y^l}{l!}\sum_{k\geq0}\frac{z^k}{k!}\args{\cos^2\argp{\frac{\pi}{2}\AdA}\delta_{\mathrm{odd}\,(k-l)}+\sin^2\argp{\frac{\pi}{2}\AdA}\delta_{\mathrm{even}\,(k-l)}}a^{\dagger \,l}\vdots(\AdA)^{k}\vdots\,a^l}\Bigg].
		\end{align}
		
		Now, employing the normal and antinormal ordering techniques as in the earlier derivation for Eq.~\eqref{eq:psucc_1dv} we obtain,
		\begin{align}
			&\,\overline{p}_{\mathrm{succ}}(00)=\notag\\
			&\frac{1}{4}\args{
				1+p+\frac{1-p+2p^2}{2G-1}\tr{\SL \argp{\frac{1-2\mu G}{2G-1}}^{\AdA}}-2p\argp{\tr{\SL \sin^2\!\left(\frac{\pi}{2}\AdA\right)}+\frac{\tr{\SL \sin^2\argp{\frac{\pi}{2}\AdA}\argp{\frac{1-2\mu G}{2G-1}}^{\AdA}}}{2G-1}}
			},
		\end{align}
		and similarly,
		\begin{align}
			&\,\overline{p}_{\mathrm{succ}}(11)=\notag\\
			&\frac{1}{4}\args{
				1+p+\frac{1-p+2p^2}{2G-1}\tr{\SL \argp{\frac{1-2\mu G}{2G-1}}^{\AdA}}-2p\argp{\tr{\SL \cos^2\!\left(\frac{\pi}{2}\AdA\right)}+\frac{\tr{\SL \cos^2\argp{\frac{\pi}{2}\AdA}\argp{\frac{1-2\mu G}{2G-1}}^{\AdA}}}{2G-1}}
			}.
		\end{align}
		The success rate of accepting both outcomes becomes,
		\begin{equation}
			\,\overline{p}_{\mathrm{succ}}=\frac{1}{2}+\frac{1-2p(1-p)}{2(2G-1)}\tr{\SL\,\argp{\frac{1-2\mu\,G}{2G-1}}^{\AdA}},
		\end{equation}
		which coincides with Eq.~\eqref{eq:psucc_1dv} for $p=0$.
		
		\section{Discrete variable teleportation fidelity with noisy resource}
		\label{app:DV_tele_fid}
		Here we derive the impact of composite amplitude and phase damping DV noise on the resource entangled state, $\ket{\Phi^+}$ required in the standard DV teleportation protocol. It is one of the four Bell states,
		$\ket{\Phi^\pm}=({\ket{00}\pm\ket{11}})/{\sqrt{2}},$ and $\ket{\Psi^\pm}=({\ket{01}\pm\ket{10}})/{\sqrt{2}}.$
		
		In the first step of the teleportation, a Bell state measurement is performed jointly on the input and one-half of the noisy resource of Eq.~\eqref{eq:noisy-Bell}. 
		The measurement is assumed to be a noiseless positive operator-valued measure (POVM), $\mathcal{M}:=\argc{\ketbra{\Phi^+}{\Phi^+},\ketbra{\Phi^-}{\Phi^-},\ketbra{\Psi^+}{\Psi^+},\ketbra{\Psi^-}{\Psi^-}}$, composed of pure Bell states.
		We are generous with the DV measurement noise for a fairly conservative comparison with our CV-DV hybrid protocol.
		With these preliminary notation the four unnormalized output states, for the DV input $\rho$, are given by,
		\begin{align}
			\tilde{\rho}^\prime(\Phi^\pm)=&\frac{1}{4}\begin{pmatrix}
				\rho_{0,0}(1+p^2)+\rho_{1,1}p(1-p)&\pm\rho_{0,1}(1-p)^2\\
				\pm\rho_{1,0}(1-p)^2&\rho_{1,1}(1-p)^2+\rho_{0,0}p(1-p)
			\end{pmatrix},
		\end{align}
		and
		\begin{align}
			\tilde{\rho}^\prime(\Psi^\pm)=&\frac{1}{4}\begin{pmatrix}
				\rho_{0,0}p(1-p)+\rho_{1,1}(1+p^2)&\pm\rho_{1,0}(1-p)^2\\
				\pm\rho_{0,1}(1-p)^2&\rho_{1,1}p(1-p)+\rho_{0,0}(1-p)^2
			\end{pmatrix}.
		\end{align}
		
		The perfect Pauli corrections $\argc{I,\sigma_z,\sigma_x, \sigma_z\sigma_x}$ are performed on the second half of the noisy resource state corresponding to the four outcomes above, so that the (normalized) teleported state is given by,
		\begin{align}
			\rho_{\mathrm{out}}=&\tilde{\rho}^\prime(\Phi^+)+\sigma_z\tilde{\rho}^\prime(\Phi^-)\sigma_z+\sigma_x\tilde{\rho}^\prime(\Psi^+)\sigma_x+\sigma_z\sigma_x\tilde{\rho}^\prime(\Psi^-)\sigma_x\sigma_z.
		\end{align}
		The fidelity of this mixed state to the pure input is then simply,
		\begin{align}
			\tr{\rho_{\mathrm{in}}\rho_{\mathrm{out}}}=&(1-p)^2+p(1-p)\rho_{0,0}\rho_{1,1}+p(\rho_{0,0}^2+\rho_{1,1}^2)\notag\\
			=&1-p+p^2-2p^2\rho_{0,0}(1-\rho_{0,0}),
		\end{align}
		which is averaged to
		\begin{align}
			\overline{\mathcal{F}}_\mathrm{tele}=&1-p+p^2-2p^2\int_0^1\D x\, x(1-x)\notag\\
			=&1-p+2p^2/3.
		\end{align}
		The above result matches the one alternatively obtained using Ref.~\cite{Horodecki1996:Teleportation}.

		\section{Noise sources}
		\label{app:noise-sources}
		Bosonic (\textsc{b}) loss, quantum-limited amplification, thermal noise, Gaussian displacement noise~(GDN) of rates $\eta$, and qubit~(\textsc{q}) depolarizing and damping channels of strengths $\eta^\prime$ and $p$ respectively, are given by
		\begin{align}
			\mathcal{N}_\mathrm{loss}[\rhob]
			=&\,\sum^\infty_{l=0}\dfrac{\mu^l}{l!}(1-\mu)^{\frac{a^\dag a}{2}}\,a^l\,\rhob\,a^{\dag\,l}\,(1-\mu)^{\frac{a^\dag a}{2}}\,,\notag\\
			\mathcal{N}_\mathrm{amp}[\rhob]
			=&\,\dfrac{1}{G}\sum^\infty_{l=0}\dfrac{(1-G^{-1})^l}{l!}\,a^{\dag l}\,G^{-\frac{a^\dag a}{2}}\,\rhob\,G^{-\frac{a^\dag a}{2}}\,a^{l}\,,\notag\\
			\mathcal{N}_\mathrm{therm}[\rhob]=&\,\mathcal{N}_\mathrm{amp}[\mathcal{N}_\mathrm{loss}[\rhob]]\left(\substack{\displaystyle G:=1+\eta\,\bar{n}\\ \displaystyle\mu:=1-\frac{1-\eta}{1+\eta\,\bar{n}}}\right),\notag\\
			\mathcal{N}_\mathrm{GDN}[\rhob]=&\,\mathcal{N}_\mathrm{amp}[\mathcal{N}_\mathrm{loss}[\rhob]]\left(\substack{\displaystyle G:=\frac{1}{1-\eta}\\ \displaystyle\mu:=\eta}\right),\notag\\
			\mathcal{N}_\mathrm{dep}[\rhoq]
			=&\,(1-\etad)\,\rhoq+\dfrac{\etad}{3}\sum^3_{j=1}\sigma_j\,\rhoq\,\sigma_j\,\,\notag\\
			\mathcal{N}_\mathrm{damp}[\rhoq]=&K_0\rhoq K_0^\dagger+K_1\rhoq K_1^\dagger\,.
			\label{eq:noise}
		\end{align}
		where $\bar{n}$ is the mean excitation and $K_0:=\ketbra{0}{0}+\sqrt{1-p}\ketbra{1}{1}$ with $K_{1}:=\sqrt{p}\ketbra{1}{1}$ for the phase damping and $\sqrt{p}\ketbra{0}{1}$ for the amplitude damping. For GDN, the specific values of $G$ and $\mu$ shown above implement a Gaussian displacement noise with variance $\sigma^2=\frac{\eta}{1-\eta}$.
		
		\section{Normal and antinormal forms}
		\label{app:ordering-forms}
		
		The antinormally ordered forms of functions of the number operator are obtained using
		\begin{align}
			\vdots F(\AdA)\vdots\ket{m}=&\sum_{n=0}^\infty c_n a^n a^{\dagger n} \ket{m}\notag\\
			=&\ket{m}\sum_{n=0}^\infty c_n \underbrace{\frac{(m+n)!}{m!}}
			_{\mathclap{\footnotesize{=(-1)^n\left(\frac{\D}{\D x}\right)^n x^{-(m+1)} \bigg|_{x=1}}}}\notag\\
			=&\sum_{n=0}^\infty c_n\left(-\frac{\D}{\D x}\right)^n x^{-(\AdA+1)}\bigg|_{x=1}\ket{m}.\notag\\
			\implies\vdots F(\AdA)\vdots=&F\left(-\frac{\D}{\D x}\right)\,x^{-(\AdA+1)}\bigg|_{x=1}.
		\end{align}
		Similar calculations for the normally ordered forms give
		\begin{align}
			:F(\AdA):\,=&F\left(\frac{\D}{\D x}\right)\,x^{\AdA}\bigg|_{x=1}.
		\end{align}
		These lead to the well-known normally and antinormally forms of the exponentials of the number operators
		\begin{align}
			\E{\lambda \AdA}=&:\E{\AdA(\E{\lambda}-1)}:\,=\E{-\lambda}\vdots\E{(1-\E{-\lambda})\AdA}\vdots
		\end{align}
		and their rearrangements
		\begin{align}
			:\E{\lambda\AdA}:\,=(1+\lambda)^{\AdA},\qquad\vdots\E{\lambda\AdA}\vdots=(1-\lambda)^{-\AdA-1}.
		\end{align}
		
		\section{Composite amplitude and phase damping noise}
		
		In this section, we derive the composition of amplitude and phase damping noise of equal noise parameters $p$.
		The DV amplitude damping channel is rewritten as
		\begin{align}
			&\mathcal{N}_{\text{damp}}^{\text{(amp)}}[\rhoq]=K_0\rho_{\textsc{q}} K_0^\dagger + K_1^{(\text{amp})}\rho_{\textsc{q}} K_1^{\dagger({\text{amp}})}\notag\\
			&=(\ketbra{0}{0}+ \sqrt{1-p}\ketbra{1}{1})\rho_{\textsc{q}}(\ketbra{0}{0}+ \sqrt{1-p}\ketbra{1}{1})+p\ketbra{0}{1}\rho_{\textsc{q}}\ketbra{1}{0}\notag\\
			&=\frac{1}{2}\bigg\{
			(1-\frac{p}{2}+\sqrt{1-p})\rhoq+(1-\frac{p}{2}-\sqrt{1-p})\sigma_3\rhoq\sigma_3+\frac{p}{2}\bigg[
			\{\rhoq,\sigma_3\}+\sigma_1\rhoq\sigma_1+\sigma_2\rhoq\sigma_2+\I\bigg(\sigma_2\rhoq\sigma_1-\sigma_1\rhoq\sigma_2\bigg)
			\bigg]
			\bigg\}.
		\end{align}
		
		Similarly, the DV phase damping channel is given by
		\begin{align}
			&\mathcal{N}_{\text{damp}}^{\text{(ph)}}[\rhoq]=K_0\rho_{\textsc{q}} K_0^\dagger + K_1^{(\text{ph})}\rho_{\textsc{q}} K_1^{\dagger({\text{ph}})}\notag\\
			&=(\ketbra{0}{0}+ \sqrt{1-p}\ketbra{1}{1})\rho_{\textsc{q}}(\ketbra{0}{0}+ \sqrt{1-p}\ketbra{1}{1})+p\ketbra{0}{0}\rho_{\textsc{q}}\ketbra{0}{0}\notag\\
			&=\left(\frac{1}{2}+\frac{\sqrt{1-p}}{2}\right)\rhoq+\left(\frac{1}{2}-\frac{\sqrt{1-p}}{2}\right)\sigma_3\rhoq\sigma_3.
		\end{align}
		
		The composite DV amplitude and phase damping channel of equal noise strength is then given by
		\begin{align}
			&\mathcal{N}_{\text{damp}}[\rhoq]:=\mathcal{N}_{\text{damp}}^{\text{(ph)}}\circ\mathcal{N}_{\text{damp}}^{\text{(amp)}}[\rhoq]=\frac{1}{4} \big\{
			c_+\rhoq+c_-\sigma_3\rhoq\sigma_3+p\big[
			\sigma_1\rhoq\sigma_1+\sigma_2\rhoq\sigma_2+\{\rhoq,\sigma_3\}+\I(\sigma_2\rhoq\sigma_1-\sigma_1\rhoq\sigma_2)
			\big]
			\big\}, 
			\label{eq:damp}
		\end{align}
		in a single parameter expression
		where 
		\begin{align}
			c_{\pm}:=&(1+\sqrt{1-p})\left(1-\frac{p}{2}\pm\sqrt{1-p}\right)+(1-\sqrt{1-p})\left(1-\frac{p}{2}\mp\sqrt{1-p}\right).
		\end{align}
		The two damping channels commute, so the ordering is irrelevant.

		\section{Averaging over random states}
		\label{app:Haar_average}
		
		The moments of pure Haar random states are given by~\cite{Harrow2013:Church, Mele2024:Introduction}
		\begin{align}
			\AVG{U\sim \mu_{\textsc{h}}}{U^{\otimes t}\ketbra{\text{\o}^{\otimes t}}{{\text{\o}}^{\otimes t}}U^{\dagger \otimes t}}=\frac{1}{{{d+t-1}\choose{t}}} P_{\mathrm{sym}}=\frac{t!}{d^{\bar{t}}}P_{\mathrm{sym}},
		\end{align}
		where $\ket{{\text{\o}}}$ is an irrelevant pure fiducial symmetric state, where $P_{\mathrm{sym}}=\frac{1}{t!}\sum_{\pi \in S_t}P_d(\pi)$ is the projector onto the space $S_t$ of symmetric permutations $\pi$ of $t$ systems of dimension $d$, $P_d(\pi)$ is the corresponding representation of the permutation on $(\mathbb{C}^d)^{\otimes t}$ and $d^{\bar{t}}=d(d+1)\cdots(d+t-1)$ are the rising factorials.
		
		Therefore, 
		\begin{align}
			&\AVG{U\sim\mu_{\textsc{h}}}{\tr{U^{ \otimes t}\ketbra{\text{\o}^{\otimes t}}{\text{\o}^{\otimes t}}U^{\dagger \otimes t}(M_1\otimes M_2\otimes \cdots \otimes M_t)}}\notag\\
			&=\frac{1}{d^{\bar{t}}}\sum_{\pi \in S_t}
			\tr{P_d(\pi)^\dagger(M_1\otimes M_2\otimes \cdots \otimes M_t)}\notag\\
			&=\frac{1}{d^{\bar{t}}} \sum_{\pi \in S_t}\prod_{c \in \operatorname{cycles}(\pi)}\mathop{\mathrm{tr}}\big\{\prod_{j \in c} M_j\big\}.
		\end{align}
		
		We assume the computational logical states to be orthogonal in the bosonic codespace defined by the normalized identity $\SL:=\big(\ketbra{0_\textsc{l}}{0_\textsc{l}}+\ketbra{1_\textsc{l}}{1_\textsc{l}}\big)/2$ and that
		\begin{align}
			M_j:=&\begin{pmatrix}
				\opinner{0_\textsc{L}}{X_j}{0_\textsc{L}} & \opinner{0_\textsc{L}}{X_j}{1_\textsc{L}}\\
				\opinner{1_\textsc{L}}{X_j}{0_\textsc{L}} & \opinner{1_\textsc{L}}{X_j}{1_\textsc{L}}
			\end{pmatrix},
		\end{align}
		with $X_j$ being the bosonic operators. 
		
		Therefore, applying the previous results to three cases, $t = 1,2,3$, and defining the logical $\rho_{\textsc{l}}:=U^\dagger\ket{{\text{\o}}}\!\!\bra{{\text{\o}}}U$, we get
		
		\begin{align}
			\AVG{\rho_L}{\Tr{\rho_{\textsc{l}} M_1}}&=\frac{1}{2}\Tr{M_1},\notag\\
			\AVG{\rho_{\textsc{l}}}{\Tr{\rho_{\textsc{l}} M_1 \rho_{\textsc{l}} M_2}}&=\AVG{\rho_{\textsc{l}}}{\Tr{\rho_{\textsc{l}}M_1}\Tr{\rho_{\textsc{l}}M_2}}=\frac{1}{6}\left(\Tr{M_1M_2}+\Tr{M_1}\Tr{M_2}\right),\notag\\
			\AVG{\rho_{\textsc{l}}}{\Tr{\rho_{\textsc{l}}M_1\rho_{\textsc{l}}M_2}\Tr{\rho_{\textsc{l}}M_2}}&=\AVG{\rho_{\textsc{l}}}{\Tr{\rho_{\textsc{l}}M_1\rho_{\textsc{l}}M_2\rho_{\textsc{l}}M_3}}\notag\\
			&=\frac{1}{12}\big(\Tr{M_1}\Tr{M_2}\Tr{M_3}-\Tr{M_3[M_1,M_2]}+2\Tr{M_1M_2M_3}\big)
		\end{align}

		With the identities and definitions set up, we have
		$\Tr{M_1}=2\tr{X_1\SL}$, $\Tr{M_1M_2}=4\tr{X_1\SL X_2 \SL}$, and $\Tr{M_1M_2M_3}=8\tr{X_1\SL X_2\SL X_3\SL}$, which leads to
		\begin{align}
			\AVG{\rho_{\textsc{l}}}{\Tr{\rho_{\textsc{l}} M_1}}&=\tr{X_1\SL}\notag\\
			\AVG{\rho_{\textsc{l}}}{\Tr{\rho_{\textsc{l}} M_1\rho_{\textsc{l}} M_2}}&=\frac{2}{3}\argp{\tr{X_1\SL X_2\SL}+\tr{X_1\SL}\tr{X_2\SL}}\notag\\
			\AVG{\rho_{\textsc{l}}}{\Tr{\rho_{\textsc{l}} M_1\rho_{\textsc{l}} M_2\rho_{\textsc{l}} M_3}}
			&=\frac{2}{3}\bigg(\tr{X_1\SL} \tr{X_2\SL}\tr{X_3\SL}-\tr{X_3\SL[X_1\SL,X_2\SL]}
			+2\tr{X_1\SL X_2\SL X_3\SL}\bigg).
		\end{align}

	\end{document}